\documentclass{article}

\usepackage{arxiv}

\usepackage[utf8]{inputenc} 
\usepackage[T1]{fontenc}    
\usepackage{hyperref}       
\usepackage{url}            
\usepackage{booktabs}       
\usepackage{amsfonts}       
\usepackage{nicefrac}       
\usepackage{microtype}      
\usepackage{lipsum}		
\usepackage{graphicx}
\usepackage{doi}
\usepackage{color,soul,comment}

\makeatletter
\AtBeginDocument{\let\hl\@firstofone}
\makeatother

\title{Course-Skill Atlas: A national longitudinal dataset of skills taught in U.S. higher education curricula}

\author{ 
\href{https://orcid.org/0000-0001-9459-2411}{\includegraphics[scale=0.06]{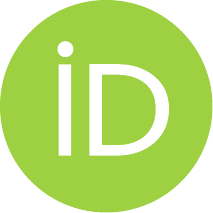}\hspace{1mm}Alireza {Javadian Sabet}}$^1$ \\
	\texttt{alj112@pitt.edu.edu} \\
	\And
	\href{https://orcid.org/0000-0002-6761-4862}{\includegraphics[scale=0.06]{orcid.pdf}\hspace{1mm}Sarah H. Bana}$^{2,3}$ \\
	\texttt{sarah.bana@gmail.com} \\
	\And
	\href{https://orcid.org/0000-0002-2375-3537}{\includegraphics[scale=0.06]{orcid.pdf}\hspace{1mm}Renzhe Yu}$^{4,5}$  \\
	\texttt{renzheyu@tc.columbia.edu} \\
	\And
	\href{https://orcid.org/0000-0001-9487-9359}{\includegraphics[scale=0.06]{orcid.pdf}\hspace{1mm}Morgan R. Frank}$^{1,3,6}$\thanks{corresponding author: Morgan R. Frank (mrfrank@pitt.edu)}  \\
	\texttt{mrfrank@pitt.edu} \\ \\
$^1$Department of Informatics and Networked Systems, University of Pittsburgh, Pittsburgh, PA 15216, USA\\
$^2$Argyros School of Business and Economics, Chapman University, Orange, CA, USA\\
$^3$Digital Economy Lab, Institute for Human-Centered Artificial Intelligence, \\Stanford University, Stanford, CA 94305, USA\\
$^4$Teachers College, Columbia University, New York, NY 10027, USA\\
$^5$Data Science Institute, Columbia University, New York, NY 10027, USA\\
$^6$Media Laboratory, Massachusetts Institute of Technology, Cambridge, MA 02139, USA
}

\renewcommand{\shorttitle}{Course-Skill Atlas}

\hypersetup{
pdftitle={A national longitudinal dataset of skills taught in U.S. higher education curricula},
pdfsubject={cs.CL},
pdfauthor={Alireza {Javadian Sabet}, Sarah H.~Bana, Renzhe Yu, Morgan R.~Frank},
pdfkeywords={Skill, Higher~Education, Future~of~Work, Labor~Economics, Complexity, ONET, Workplace~Activity}
}

\begin{document}
\maketitle

\begin{abstract}
Higher education plays a critical role in driving an innovative economy by equipping students with knowledge and skills demanded by the workforce.
While researchers and practitioners have developed data systems to track detailed occupational skills, such as those established by the U.S. Department of Labor (DOL), much less effort has been made to document which of these skills are being developed in higher education at a similar granularity.
Here, we fill this gap by presenting Course-Skill Atlas -- a longitudinal dataset of skills inferred from over three million course syllabi taught at nearly three thousand U.S. higher education institutions.
\hl{To construct Course-Skill Atlas, we apply natural language processing to quantify the alignment between course syllabi and detailed workplace activities (DWAs) used by the DOL to describe occupations. 
We then aggregate these alignment scores to create skill profiles for institutions and academic majors. 
Our dataset offers a large-scale representation of college education's role in preparing students for the labor market.}
Overall, Course-Skill Atlas can enable new research on the source of skills in the context of workforce development and provide actionable insights for shaping the future of higher education to meet evolving labor demands, especially in the face of new technologies.

\end{abstract}


\section*{\hl{Background \& Summary}}

Skills are essential components of jobs and shape the career outcomes of workers in the labor market. 
Therefore, systematically studying skills and their sources is essential for predicting workers' career trajectories and macro-level workforce dynamics~\cite{deming2018skill, doi:10.1177/0730888490017004002, doi:10.1177/0730888490017004003, warhurst2017oxford}. 
For example, recent research finds increasing demand for social skills for modern, flexible team-based work environments based on required skills in job postings~\cite{deming2017value}. 
In response to shifts in skills, employers need to consider skill profiles and skill development in their hiring and training. 
For instance, employers subjectively perceive the skill content of college majors when determining the requirements to include in online job advertisements~\cite{hemelt2023college}.
Combined, the focus on skills in the labor market warrants a similar perspective on the sources of skills during workforce development and talent acquisition.

Higher education is arguably the most important source of skill development, which facilitates both economic and social mobility~\cite{haveman2006role}. 
In the past few decades, empirical studies have consistently demonstrated that college-educated individuals earn higher wages, achieve more extensive professional networks, and collectively experience greater inter-generational upward mobility~\cite{chetty2017mobility,kerckhoff2001education}. 
Non-college educated workers now engage more in less skilled tasks than their counterparts compared to previous eras~\cite{acemoglu2011skills} and tend towards low-wage occupations. 
On the other hand, the economic returns of higher education vary across fields of study due to differing skill sets imparted by college majors~\cite{altonji2016analysis, altonji2012heterogeneity}. 
They also vary because of institutional selectivity~\cite{triventi2013role}. 
Moreover, students from different demographic and socioeconomic backgrounds are sorted into different educational trajectories due to existing structural inequalities which may hinder the social mobility that higher education is intended to foster~\cite{LOVENHEIM2023187}. In recent years, as elevated dropout rates~\cite{dropout} and rising unemployment or underemployment rates of college graduates fuel concerns around the efficacy of higher education~\cite{bonvillian2021workforce, taylor2011college}, it is important to better understand \textit{how} higher education imparts skills and prepares students for the labor market. 
This will require moving beyond using degree or credit attainment as proxies for skills, as these proxies fail to capture subtle variations in educational experiences between students in the same major or institution.

\hl{Recently, large-scale data about curricular and job \textit{content} has become available in digital formats, which provides a new possibility of examining the mechanism of skill development. 
Some researchers have started to leverage these new data sources to connect higher education and jobs using natural language processing techniques{~\cite{Yu2021,light2024student,borner2018skill,Garcia2024,Chau2023,Desikan2022,Chang2022,Lastra2021}}.
For example, B\"{o}rner and colleagues{~\cite{borner2018skill}} provided one of the earliest large-scale analysis of the alignment between college courses, job vacancies, and academic research. They leveraged an established skill taxonomy and connected the three pieces via skill mentions in their content. More recently, Light{~\cite{light2024student}} measures changes in university course offerings over time by quantifying the semantic overlap between course descriptions and job postings. 
This novel line of research has provided emerging evidence of mismatches between labor market demands and skills taught in courses, as well as the uneven distribution of these mismatches along such dimensions as major areas, institutions, and geographical locations.}

\hl{Despite the important findings, almost all the empirical studies to date acquire educational and job content information through proprietary data contracts with private vendors, and few have released their derived data about education-occupation alignment, making replication and extension efforts challenging for other researchers.}
In this context, we provide a new algorithmic pipeline and a public dataset to help analyze the skill development in American higher education~\cite{ourDataset}. 
We first introduce \texttt{Syllabus2O\textsuperscript{*}NET}, a natural language processing (NLP) framework designed to identify and interpret skills from curricular content, in line with the O\textsuperscript{*}NET taxonomy~\cite{ONET} used by the U.S. Department of Labor (DOL) (see Figure~\ref{fig:workflow}). 
Applying \texttt{Syllabus2O\textsuperscript{*}NET} to the most extensive dataset of university course syllabi, we then present Course-Skill Atlas --- a longitudinal, national dataset of inferred skill profiles across different institutions, academic majors, and student populations in the United States. 
To validate this dataset, we perform qualitative and quantitative explorations of the identified skills in reference to existing studies. 
We further discuss a handful of potential use cases of Course-Skill Atlas, including quantifying skill-salary correlations, analyzing temporal trends in curriculum design, and revealing gender skills gaps based on major and institution.


Overall, our provision makes three intellectual and practical contributions.
First, we present a computational framework to describe the content alignment between education and workforce. While we developed the methodology based on two specific document types, it is applicable to other types of documents as well. 
Second, we provide an essential, public data source on the granular skill profiles of institutions, which can facilitate future research in such fields as higher education, labor economics, and future of work, especially in an era marked by rapid technological advancements and shifting economic landscapes~\cite{akour2022higher, brasca2022technology}. 
Third, our validation analyses illustrate some macro-level patterns of skills taught in higher education that warrant more in-depth research in the future.

\section*{Methods}
\label{sec:methods}

\subsection*{Materials}
\label{sec:materials}

\subsubsection*{Open Syllabus Project Dataset}
\label{sec:ospdataset}

Open Syllabus Project (OSP) ({\url{https://opensyllabus.org/}}) is a non-profit organization that curates a vast archive of over $20.9$ million course syllabi from higher education institutions worldwide. 
The organization aims to map and analyze the curriculum across thousands of institutions, providing insights into the most commonly taught texts and subjects.
\hl{OSP's syllabi data comes from ({\romannumeral 1}) scraped content from universities' syllabi repositories, ({\romannumeral 2}) a broad web crawler with seeds from CommonCrawl (}\url{https://commoncrawl.org/}\hl{) and manual curation, ({\romannumeral 3}) Internet Archive for 2021 Open Syllabus crawl (}\url{https://archive.org/details/OPENSYLLABUS-20210506220126-crawl804}\hl{), and ({\romannumeral 4}) syllabi donation from institutions and individuals.
Through a research contract, we analyze one version 2.1 of the OSP data, which encompasses nearly 8 million course syllabi worldwide among which $3,162,747$ syllabi across $62$ fields of study (FOS) belong to $2,761$ colleges and universities in the United States.}
Each course syllabus contains features such as course description, language, year, field of study, and information about the institution.
In this paper, `major' and `FOS' are used interchangeably.

\subsubsection*{O\textsuperscript{*}NET}
\label{sec:onetgeneral}

O\textsuperscript{*}NET (Occupational Information Network) (\url{https://www.onetonline.org/}) stands as a comprehensive database detailing worker attributes and job characteristics. 
Developed under the sponsorship of the U.S. Department of Labor/Employment and Training Administration, O\textsuperscript{*}NET is essential for in-depth labor market and workforce analyses~\cite{lim2023location, frank2018small, chau2023connecting, alabdulkareem2018unpacking, frank2019toward, moro2021universal, frank2024network, agnihotri2024managerial, cabell2023supporting}.
Educators, career counselors, and workforce development professionals leverage O\textsuperscript{*}NET for evaluating job requirements against worker qualifications, aiding in curriculum development, career guidance, and labor market analysis~\cite{national2010database}. 
Occupations form the core of the O\textsuperscript{*}NET system, around which a standardized hierarchical taxonomy is organized, allowing for detailed analysis and comparison across diverse professional roles including:

\begin{itemize}
    \item \textbf{Worker Characteristics}: These are essential in understanding the potential and capacity of the workforce. We specifically focus on:
        \begin{itemize}
        \item \textit{Ability} (\url{https://www.onetonline.org/find/descriptor/browse/1.A}): The performance of individual workers is influenced by their enduring abilities, the most granular component of worker characteristics. There are $52$ abilities categorized into four key areas: cognitive, physical, psychomotor, and sensory.
        \end{itemize} 
    \item \textbf{Occupational Requirements}: These requirements define the specific demands of jobs and are integral to occupational definitions within the taxonomy. We analyze:
        \begin{itemize}
            \item \textit{Detailed Work Activity (DWA)} (\url{https://www.onetcenter.org/dictionary/20.1/excel/dwa_reference.html}): DWAs are precise descriptions of tasks and responsibilities of specific jobs. There are more than $2,000$ DWAs across different occupations, which help understand the day-to-day activities and skills required for a particular role.
            \item \textit{Task Statement} (\url{https://www.onetcenter.org/dictionary/20.1/excel/task_statements.html}): Task is the basic unit of work. There are nearly $18,000$ tasks in total, which provide the most detailed overview of job responsibilities.
    \end{itemize}

\end{itemize}

Through its multidimensional taxonomy centered around occupations, O\textsuperscript{*}NET not only facilitates a detailed understanding of job roles but also significantly aids in bridging educational preparation with labor market demands. 
This structure makes it an invaluable tool for policymakers, educators, and employment specialists. 
By focusing on occupations, O\textsuperscript{*}NET enables a targeted analysis of the workforce, enhancing the relevance and applicability of labor market data in various professional settings.

\subsection*{Skill Inference Framework (\texttt{Syllabus2O\textsuperscript{*}NET})}
\label{sec:ospSkillsExtraction}

\begin{figure}
    \centering
    \includegraphics[trim=0 15 0 0,clip,width=1\textwidth]{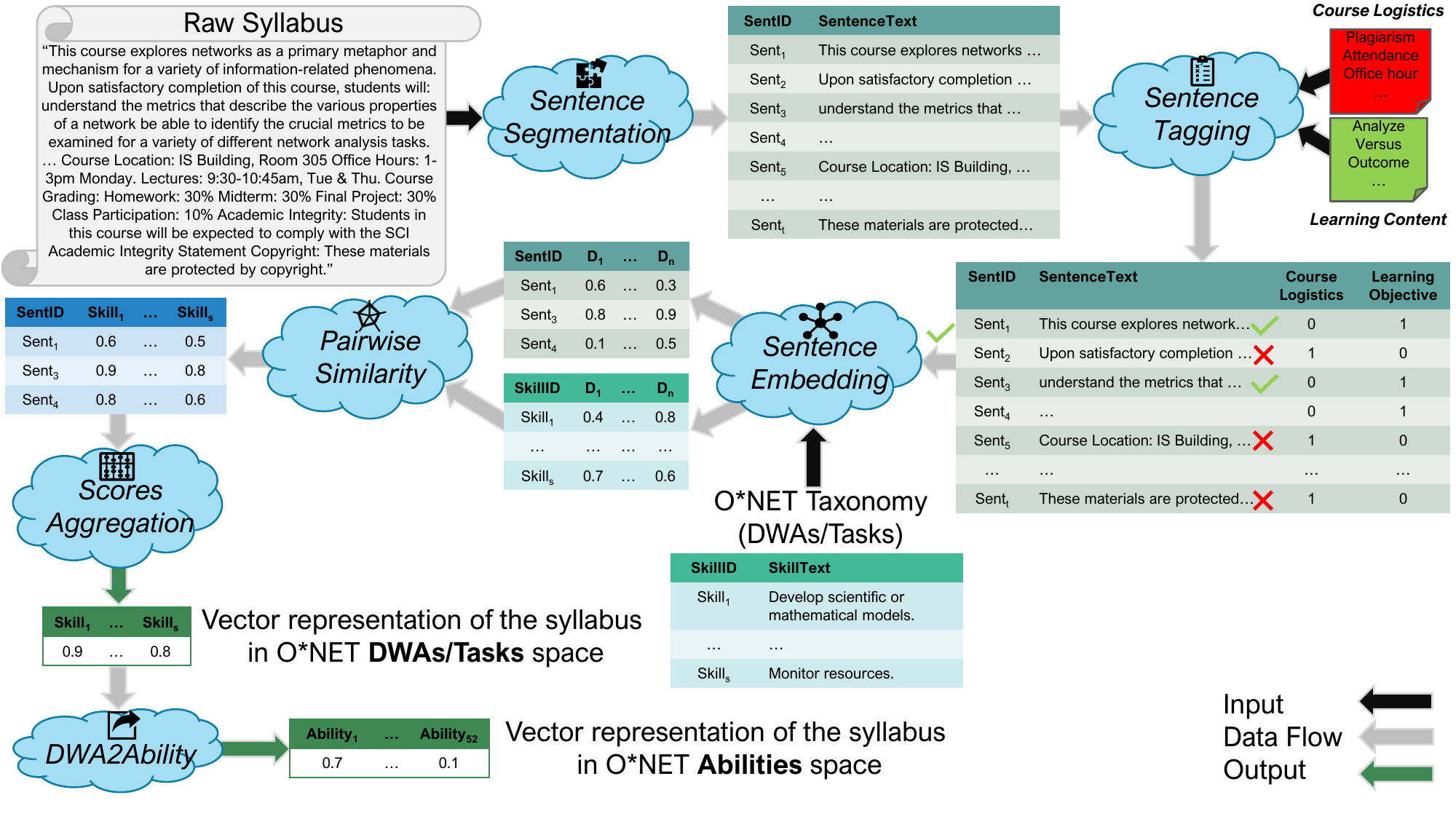}
    \caption{
    \textbf{\hl{The \texttt{Syllabus2O\textsuperscript{*}NET} skill inference framework.}} This natural language processing framework converts a course syllabus into a vector representing its coverage of individual ``skills'' defined by O\textsuperscript{*}NET Detailed Work Activity (DWA) or Task. The pipeline receives a course syllabus as input and segments the raw texts into individual sentences. Then, using a curated dictionary, it identifies and keeps sentences related to learning content and transforms each sentence into a high-dimensional vector with sentence embedding (SBERT). Meanwhile, each skill is vectorized with the same approach. Next, pairwise cosine similarities between the embeddings of skills and learning content sentences are computed. Then, each skill's maximum similarity score across the learning content sentences is used to indicate the skill coverage of the syllabus. \hl{Finally, \texttt{DWA2Ability}, which is composed of 52 Random Forest Regressors, maps the inferred skill coverage to related worker abilities.}
} 
    \label{fig:workflow}
\end{figure}

Figure~\ref{fig:workflow} provides an overview of our \texttt{Syllabus2O\textsuperscript{*}NET} skill inference framework. 
This framework leverages natural language processing to estimate skill coverage in curricular content. 
\texttt{Syllabus2O\textsuperscript{*}NET} begins by taking the raw texts of a course syllabus, which typically include course logistics (e.g., scheduling and grading rubrics) and learning content (e.g., learning objectives). 

We then use Stanza~\cite{qi2020stanza} to partition the raw texts into individual sentences. 
\hl{Stanza leverages pre-trained neural network models to syntactically parse documents into sentences across diverse contexts and languages and has been shown to work even in the presence of complex punctuation and formatting{~\cite{qi2020stanza}}.} 
This tool helps extract \hl{$322,473,524$} sentences from the $3,162,747$ course syllabi in the OSP dataset. 
On average, each syllabus contains \hl{$101.96$ sentences (median $83$)}.

Because not all sections of a syllabus reflect the subject matter and covered skills of the course, a human-in-the-loop approach is deployed to remove sentences pertaining to course logistics while keeping sentences about learning content.
To do so, we compiled two distinct lists of keywords for labeling sentences, one for course logistics that includes 356 phrases (e.g., ``plagiarism,'' ``attendance,'' and ``office hours''), and another for learning content including 51 phrases  (e.g., ``analyze,'' ``versus,'' and ``outcome''). The complete lists can be found on \hl{``course\_logistics\_terms'' and ``learning\_content\_terms'' files on Figshare}~\cite{ourDataset} and on the project's GitHub page.
We removed sentences from each syllabus that contained phrases related to course logistics or that lacked language related to learning content, resulting in the removal of \hl{$85.82\%$} of the sentences in the raw data.
After this cleaning process, each syllabus on average contains \hl{$17.61$} learning content sentences \hl{(median = $12$) (see Table}~\ref{tab:fos_sent_dist} for details on the statistics of ``learning content'' sentence counts by FOS).

\begin{table}
\small
\centering
\resizebox{1\textwidth}{!}{
\begin{tabular}{lccc|lccc}
\hline
\textbf{Field of Study} & \textbf{\#Sent.} & \textbf{\# L. Sent.} & \textbf{\% L. Sent.} & \textbf{Field of Study} & \textbf{\# Sent.} & \textbf{\# L. Sent.} & \textbf{\%L. Sent.} \\ \hline
Accounting              & 101            & 15                 & 14.61\%              & Hebrew                  & 65              & 7                  & 11.32\%             \\
Agriculture             & 60             & 8                  & 15.11\%              & History                 & 88              & 9                  & 10.62\%             \\
Anthropology            & 86             & 13                 & 14.29\%              & Japanese                & 84              & 10                 & 13.12\%             \\
Architecture            & 73             & 14                 & 20.45\%              & Journalism              & 105             & 13                 & 12.22\%             \\
Astronomy               & 76             & 10                 & 12.50\%               & Law                     & 60              & 7                  & 12.12\%             \\
Atmospheric Sciences    & 66             & 11                 & 14.29\%              & Library Science         & 107             & 13                 & 12.04\%             \\
Basic Computer Skills   & 88             & 11                 & 13.21\%              & Linguistics             & 72              & 9                  & 11.43\%             \\
Basic Skills            & 76             & 11                 & 14.29\%              & Marketing               & 106             & 16                 & 15.79\%             \\
Biology                 & 85             & 12                 & 13.25\%              & Mathematics             & 75              & 10                 & 12.73\%             \\
Business                & 97             & 15                 & 15.22\%              & Mechanic/Repair Tech  & 62              & 10                 & 17.95\%             \\
Chemistry               & 83             & 11                 & 13.16\%              & Media/Communications  & 107             & 14                 & 13.73\%             \\
Chinese                 & 60             & 8                  & 12.10\%               & Medicine                & 95              & 14                 & 15.79\%             \\
Classics                & 47             & 4                  & 9.30\%                & Military Science        & 71              & 11                 & 14.35\%             \\
Computer Science        & 67             & 10                 & 14.29\%              & Music                   & 62              & 8                  & 13.04\%             \\
Cosmetology             & 85             & 12                 & 15.79\%              & Nursing                 & 95              & 16                 & 16.95\%             \\
Criminal Justice        & 79             & 12                 & 15.38\%              & Nutrition               & 85              & 12                 & 14.46\%             \\
Culinary Arts           & 61             & 14                 & 22.22\%              & Philosophy              & 76              & 9                  & 12.50\%              \\
Dance                   & 59             & 11                 & 17.54\%              & Physics                 & 61              & 8                  & 14.19\%             \\
Dentistry               & 69             & 16                 & 25.93\%              & Political Science       & 96              & 12                 & 12.23\%             \\
Earth Sciences          & 59             & 9                  & 14.29\%              & Psychology              & 105             & 16                 & 15.09\%             \\
Economics               & 74             & 11                 & 13.92\%              & Public Safety           & 60              & 10                 & 16.03\%             \\
Education               & 121            & 23                 & 19.74\%              & Religion                & 84              & 11                 & 12.61\%             \\
Engineering             & 51             & 9                  & 16.28\%              & Sign Language           & 78              & 12                 & 15.69\%             \\
English Literature      & 97             & 11                 & 10.99\%              & Social Work             & 139             & 26                 & 19.28\%             \\
Film and Photography    & 68             & 11                 & 14.88\%              & Sociology               & 100             & 14                 & 13.92\%             \\
Fine Arts               & 79             & 13                 & 15.75\%              & Spanish                 & 95              & 10                 & 10.60\%              \\
Fitness and Leisure     & 54             & 10                 & 17.19\%              & Theatre Arts            & 63              & 11                 & 17.68\%             \\
French                  & 69             & 7                  & 10.45\%              & Theology                & 113             & 16                 & 14.29\%             \\
Geography               & 70             & 10                 & 14.29\%              & Transportation          & 94              & 12                 & 13.46\%             \\
German                  & 71             & 8                  & 11.39\%              & Veterinary Medicine     & 70              & 10                 & 11.39\%             \\
Health Technician       & 75             & 11                 & 16.22\%              & Women's Studies         & 81              & 13                 & 15.19\%             \\ \hline
\end{tabular}
}
\caption{\textbf{Sentence statistics per FOS.} The table presents the median of the number of sentences (\# Sent.), number of identified Learning Objective related sentences (\# L. Sent.), and the percentage of the identified Learning Objective related sentences (\% L. Sent.) per FOS.}
\label{tab:fos_sent_dist}
\end{table}

Next, we compute the semantic similarity between each O\textsuperscript{*}NET DWA or Task (hereafter, ``skill'') and each sentence in a syllabus.
\hl{SBERT{~\cite{reimers2019sentence}}, a neural language model with a Siamese network structure, is used to convert a sentence into a fixed-size vector (also called ``embedding'') that encodes its semantic meaning.} 
\hl{We choose SBERT over alternative language models due to its diverse training corpora, faster computation, and superior performance on benchmark tasks.
SBERT is trained on a diverse range of more than 1 billion sentences including S2ORC: The Semantic Scholar Open Research Corpus{~\cite{lo-wang-2020-s2orc}}, WikiAnswers Corpus{~\cite{Fader14}}, PAQ: 65 Million Probably-Asked Questions{~\cite{lewis2021paq}}, and GooAQ: Open Question Answering with Diverse Answer Types{~\cite{gooaq2021}}.} 
Specifically, we implement the ``all-mpnet-base-v2'' model~\cite{Reimers_Inui_2019} to embed each ``learning content'' sentence in course syllabi and each skill descriptor into a 768-dimension semantic space. 
Pairwise cosine similarity between these embeddings are calculated to measure semantic similarity between learning content and skills. For instance, ``understand the metrics that describe the various properties of a network be able to identify the crucial metrics to be examined for a variety of different network analysis tasks'' (in Figure~\ref{fig:workflow}) is semantically similar to the O\textsuperscript{*}NET DWA ``develop scientific or mathematical models,'' with a cosine similarity of $0.9$.

\hl{Finally, we create a vector for each syllabus to measure how much each skill is covered in the course, based on the semantic similarities. Specifically, for a given skill, we select the maximum similarity score across all the sentences within the syllabus (i.e., the score of the most similar sentence).}
\hl{ 
This approach captures the most relevant skill information each course aims to develop and is particularly robust for handling the significant variations in syllabi's length, detail, and structures.}
With this construction, the syllabus vectors have 2,070 dimensions for DWAs and 17,992 for Tasks.

Some studies might aim at depicting the competencies of workers and do not have detailed information about jobs.
In line with this use case, we further establish the connection between learning content and O\textsuperscript{*}NET abilities by mapping inferred skills to abilities. Because O\textsuperscript{*}NET does not provide a standardized crosswalk linking DWAs, tasks, and abilities, we create a subprocess \texttt{DWA2Ability} to achieve this.
We start with the O\textsuperscript{*}NET database profiles of DWAs for each occupation.
Next, we extract importance scores of each O\textsuperscript{*}NET ability within each occupation.
We formulate a map between the two sets of occupation profiles as a regression using DWAs as independent variables and ability scores as dependent variables. 
We train a Random Forest Regressor~\cite{scikit-learn} for each ability and fine-tune hyperparameters via Grid Search and 5-fold cross-validation. 
This approach yielded 52 models (i.e., one per O\textsuperscript{*}NET ability), each achieving mean squared error of at most 0.025 (see Table~\ref{tab:abilities_mse} for details on model performance). 
Using the trained models, we map syllabi's DWA scores to abilities. 
\hl{If a syllabus does not teach content that provides the students with a certain ability, the corresponding ability score is $0$.}

\begin{table}
\centering
\small
\resizebox{1\textwidth}{!}{
\begin{tabular}{lc|lc|lc|lc}
\hline
\multicolumn{1}{l}{\textbf{Ability}} & \textbf{MSE} & \multicolumn{1}{l}{\textbf{Ability}} & \textbf{MSE} & \multicolumn{1}{l}{\textbf{Ability}} & \textbf{MSE} & \multicolumn{1}{l}{\textbf{Ability}} & \textbf{MSE} \\ \hline
Arm-Hand Steadiness                  & 0.025        & Fluency of Ideas                     & 0.014        & Number Facility                      & 0.012        & Speech Clarity                       & 0.008        \\
Auditory Attention                   & 0.014        & Glare Sensitivity                    & 0.017        & Oral Comprehension                   & 0.010        & Speech Recognition                   & 0.014        \\
Category Flexibility                 & 0.016        & Gross Body Coordination              & 0.013        & Oral Expression                      & 0.011        & Speed of Closure                     & 0.010        \\
Control Precision                    & 0.019        & Gross Body Equilibrium               & 0.018        & Originality                          & 0.014        & Speed of Limb Movement               & 0.020        \\
Deductive Reasoning                  & 0.011        & Hearing Sensitivity                  & 0.012        & Perceptual Speed                     & 0.012        & Stamina                              & 0.019        \\
Depth Perception                     & 0.014        & Inductive Reasoning                  & 0.010        & Peripheral Vision                    & 0.013        & Static Strength                      & 0.024        \\
Dynamic Flexibility                  & 0.007        & Information Ordering                 & 0.017        & Problem Sensitivity                  & 0.010        & Time Sharing                         & 0.013        \\
Dynamic Strength                     & 0.016        & Manual Dexterity                     & 0.023        & Rate Control                         & 0.019        & Trunk Strength                       & 0.017        \\
Explosive Strength                   & 0.016        & Mathematical Reasoning               & 0.011        & Reaction Time                        & 0.020        & Visual Color Discrimination          & 0.019        \\
Extent Flexibility                   & 0.019        & Memorization                         & 0.013        & Response Orientation                 & 0.012        & Visualization                        & 0.017        \\
Far Vision                           & 0.011        & Multilimb Coordination               & 0.022        & Selective Attention                  & 0.008        & Wrist-Finger Speed                   & 0.024        \\
Finger Dexterity                     & 0.015        & Near Vision                          & 0.018        & Sound Localization                   & 0.025        & Written Comprehension                & 0.012        \\
Flexibility of Closure               & 0.014        & Night Vision                         & 0.015        & Spatial Orientation                  & 0.017        & Written Expression                   & 0.014        \\ \hline
\end{tabular}
}
\caption{\textbf{\texttt{DWA2Ability} models training performance for each ability.} The mean squared error (MSE) obtained from the 5-fold cross-validation (CV) for finding the best model of ability.}
\label{tab:abilities_mse}
\end{table}

\paragraph{Negative values in cosine similarity}
\hl{To determine how much each skill is covered in a course, we calculate the cosine similarity between the skill and each sentence in a syllabus. 
Cosine similarity ranges from $-1$ to $1$, where a value of $1$ indicates that the two vectors (or sentences in this context) are perfectly aligned and have the same direction, while a value of $-1$ indicates that the vectors are diametrically opposed. 
In general, a negative cosine similarity value suggests that the two sentences are not only dissimilar but also convey opposite or contrasting meanings. 
However, in the context of skill inference, the meaning of dissimilarity might not always be applicable. 
For example, the top five DWAs with the most negative values include ``Trim trees or other vegetation,'' ``Adjust the tension of nuts or bolts,'' ``Install carpet or flooring,'' ``Install trim or paneling,'' and ``Apply sealants or other protective coatings.''
In the output of our \texttt{Syllabus2O\textsuperscript{*}NET} pipeline, fewer than 10 DWAs have negative values in more than $5\%$ of the syllabi, with the highest being $7.17\%$. For transparency and flexibility, we have kept these negative values as they are in the released aggregated datasets, allowing users to take appropriate actions according to their specific needs (e.g., change them to zero or normalize them).}

\subsection*{Skill Normalization}
\label{sec:SkillNormalization}

Some skills, such as ``Maintain student records'' and ``Document lesson plans'', are ubiquitous across fields of study (FOS) and, therefore, do not distinguish the learning content of one FOS from another.
To address this issue and control for widespread skills we propose two approaches when using our data. 
Although we use them in some of the validation analysis, the published dataset remains intact.

The first approach is applying Revealed Comparative Advantage (RCA) (a.k.a., ``location quotient''~\cite{hoen2006measurement, alabdulkareem2018unpacking, reflection, hausmann2011network, hartmann2017linking}).
\hl{RCA is a concept used to identify areas where an entity has a relatively high presence compared to others. 
In our context, RCA helps to reveal which skill $s$ most strongly distinguishes one FOS $m$ from others.}
RCA is calculated as follows:
\begin{equation}
    rca(m,s)=\frac{onet(m,s)/\displaystyle\sum_{s'\in S}onet(m,s')}{\displaystyle\sum_{m'\in S}onet(m',s)/\sum_{m'\in M,s'\in S}onet(m',s')},
    \label{eq:rca}
\end{equation}
where $m\in M$ denotes a FOS (i.e., a college major) and $s\in S$ denotes a skill (i.e., an O\textsuperscript{*}NET DWA).
If $rca(m,s)>1$, then $s$ is more related to $m$ than would be expected across all DWAs and all FOS; therefore, $s$ is a relatively distinctive skill identifying $m$ more strongly than other FOS.

The second approach, which we used to produce Fig.~\ref{fig:dwa_mean_rca_3_samples},
is to mask frequent skills in empirical analyses that use the skill scores. 
What is considered ``frequent'' should depend on the specific application, so we do not mask any skills in the published dataset.
However, as a useful resource, \hl{we create a table ``top10\_DWA\_per\_FOS'' on Figshare}~\cite{ourDataset} which \hl{lists the top 10 inferred workplace activities with the highest average DWA scores per FOS.}

\section*{Data Records}
\label{sec:datarecords}

\hl{The Course-Skill Atlas dataset is freely available at Figshare}~\cite{ourDataset}.

\subsection*{Schema and Variables}

\begin{figure}
    \centering
    \includegraphics[trim=20 5 20 5,clip,width=1\textwidth]{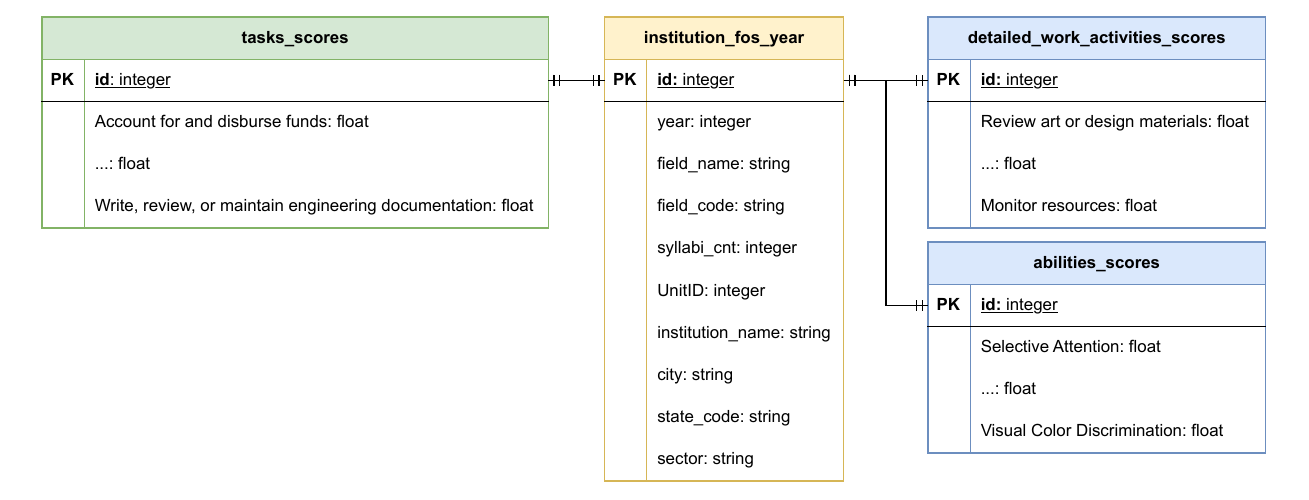}
    \caption{\hl{\textbf{The entity relationship diagram of the skills extracted from U.S. course syllabi.}} 
    In each table, PK represents the table's primary key.
    ``institution\_fos\_year'' comprise the main data table encompassing $281,153$ records.   
    For each corresponding id from ``institution\_fos\_year'' Table, ``task\_scores'', ``detailed\_work\_activities\_scores'', and ``abilities\_scores'' tables contain the scores for $17,992$ tasks, $2,070$ DWAs, and $52$ abilities respectively inferred using  \texttt{Syllabus2O\textsuperscript{*}NET}). 
    For brevity, we replaced the remaining tasks, DWAs, and abilities with ``\dots''. Lines connecting tables indicate the presence of a relational table.} 
    \label{fig:dataset_tables_schema}
\end{figure}

Our research contract with OSP requires that we do not release information at the individual syllabus level. 
As such, we create a dataset of inferred skills aggregated at the institution-year-FOS level.
\hl{The Course-Skill Atlas dataset includes three key components: \textbf{DWAs}, \textbf{Tasks}, and \textbf{Abilities}. 
Each represents the corresponding vectors produced by \texttt{Syllabus2O\textsuperscript{*}NET} and aggregated at our chosen granularity.} 
Figure{~\ref{fig:dataset_tables_schema}} provides the entity relationship diagram of these components.

Each record in these datasets includes the year, institution name, and UnitID (i.e., the unique identifier assigned by Integrated Postsecondary Education Data System (IPEDS) (\url{https://nces.ed.gov/collegenavigator/}) to each institution), the geographical location of the institution (i.e., the city and state), the FOS name along with its CIP code(s), \hl{as well as the institution's sector --- one of nine institutional categories created by combining an institution's control and level (e.g., ``public 4-year or above''), which we obtained from the Carnegie Classification of Institutions of Higher Education (CCIHE)} (\url{https://carnegieclassifications.acenet.edu/}).

\hl{The field\_code string contains one or more IPEDS CIP codes (}\url{https://nces.ed.gov/ipeds/}\hl{), representing the field(s) of study most associated with the syllabus. OSP's field classifier relies on the IPEDS 2010 CIP taxonomy (}\url{https://nces.ed.gov/ipeds/cipcode/default.aspx?y=55}\hl{) to determine the most relevant field of study (FOS) for each syllabus. 
CIP codes are structured in lengths of two, four, and six digits, where two-digit codes represent a broad discipline, four-digit codes represent subdivisions of that discipline, and six-digit codes provide further subdivisions. For instance, the two-digit CIP code '01' corresponds to ``Agriculture, Agriculture Operations, and Related Sciences''; within this category, the four-digit code ``01.01'' denotes ``Agricultural Business and Management,' and ``01.0103'' specifies ``Agricultural Economics.''
OSP’s FOS classifier is trained and tested on a curated subset of the CIP taxonomy that OSP has found most effective for describing syllabi. Occasionally, OSP combines codes, but only within the same two-digit branch of the taxonomy. In these instances, codes are separated by a forward slash (`/'). For example, the code ``45.09/45.10'' merges ``International Relations and National Security Studies'' and ``Political Science and Government,'' both of which fall under the two-digit code `45' for ``Social Sciences.'' 
This combined FOS is labeled as ``Political Science.''
Note that the the assigned CIP code(s) are consistent across all the syllabi; meaning that all ``Political Science'' syllabi are mapped to the same list of CIP codes, i.e.,  ``45.09/45.10'' (see Table field\_name\_and\_code on Figshare {~\cite{ourDataset}} for the list of fields name and their CIP code(s) mappings).
If OSP is unable to confidently assign an academic field to the syllabus, the value of this column is null.}
\hl{Moreover, we enriched each record by the  sector of the institutions (i.e., control and level combined (}\url{https://carnegieclassifications.acenet.edu/wp-content/uploads/2023/03/CCIHE2021-PublicData.xlsx}\hl{).
Further university characteristics can be added to the dataset by merging the syllabi Table and the target dataset on the UnitID variable.}

\hl{The aggregated scores are the \textbf{average scores} across all the syllabi belonging to the corresponding year, university, and FOS.
For example, let's consider a given year, university, and FOS which has two syllabi. 
The score of $DWA_{1}$ in $Syllabus_{a}$ is 0.8 and the score of the same DWA in $Syllabus_{b}$ is 0.6.
Taking the average of the two scores, the aggregated score of $DWA_{1}$ for the given triplet is 0.7.}

For the remainder of this paper, we only use DWA scores for various descriptive and validation analyses, which can be easily applied to task and ability scores as well.

\subsection*{\hl{Descriptive Statistics}}

\begin{figure}[h]
    \centering
    \includegraphics[trim=90 1200 40 0,clip,width=1\textwidth]{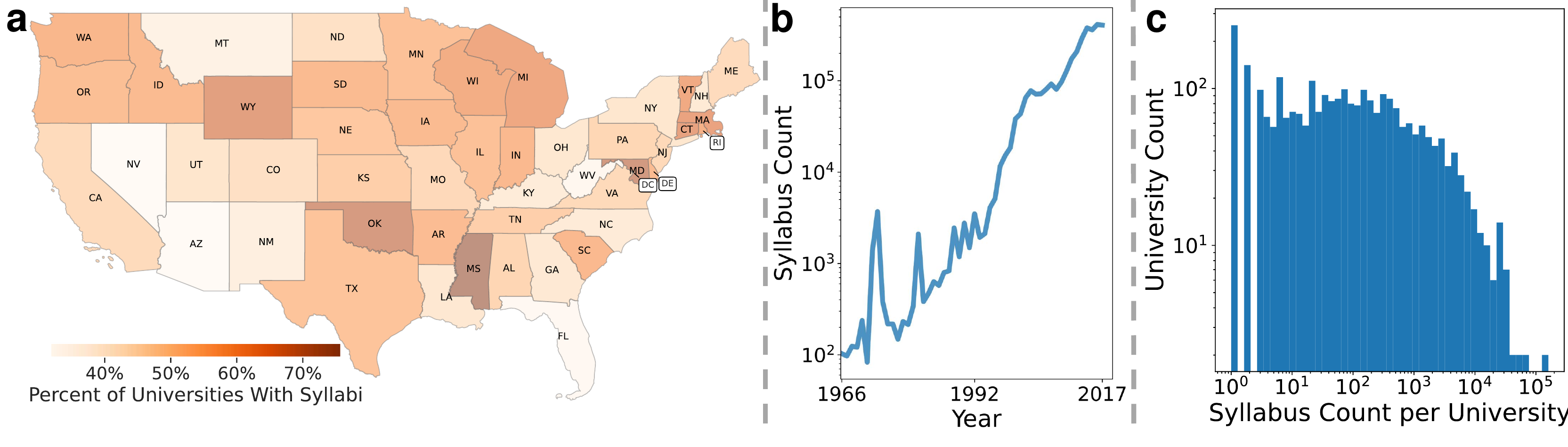}
    \caption{\textbf{Descriptive statistics of the Open Syllabus Project (OSP) dataset.} \hl{(a) Percentage of the universities with at least eight course syllabi (25\textsuperscript{th} percentile) available per state. 
    (b) Total number of syllabi per year.}
    (c) The syllabus count per university across all years and all FOS. }
    \label{fig:distributions_all_in_one}
\end{figure}

Figure~\ref{fig:distributions_all_in_one} depicts the geographical, temporal, and institutional distribution of the syllabi data.
Across each U.S. state, between $32\%$ and $76\%$ of the postsecondary institutions in each state provide at least eight course syllabi (i.e., corresponding to the 25\textsuperscript{th} percentile. See Fig.~\ref{fig:distributions_all_in_one}a).
\hl{The majority of the syllabi belong to the period post-2000, with sparse coverage between 1966 and 1999. (See Fig.{~\ref{fig:distributions_all_in_one}}b) }
Across the entire data set, the majority of universities (nearly 2000) contribute at least 10 syllabi to the data, but some contribute up to $10^5$ syllabi across all FOS (see Fig.~\ref{fig:distributions_all_in_one}c).

\begin{figure}[h!]
    \centering
    \includegraphics[trim=0 0 0 0,clip,width=0.65\textwidth]{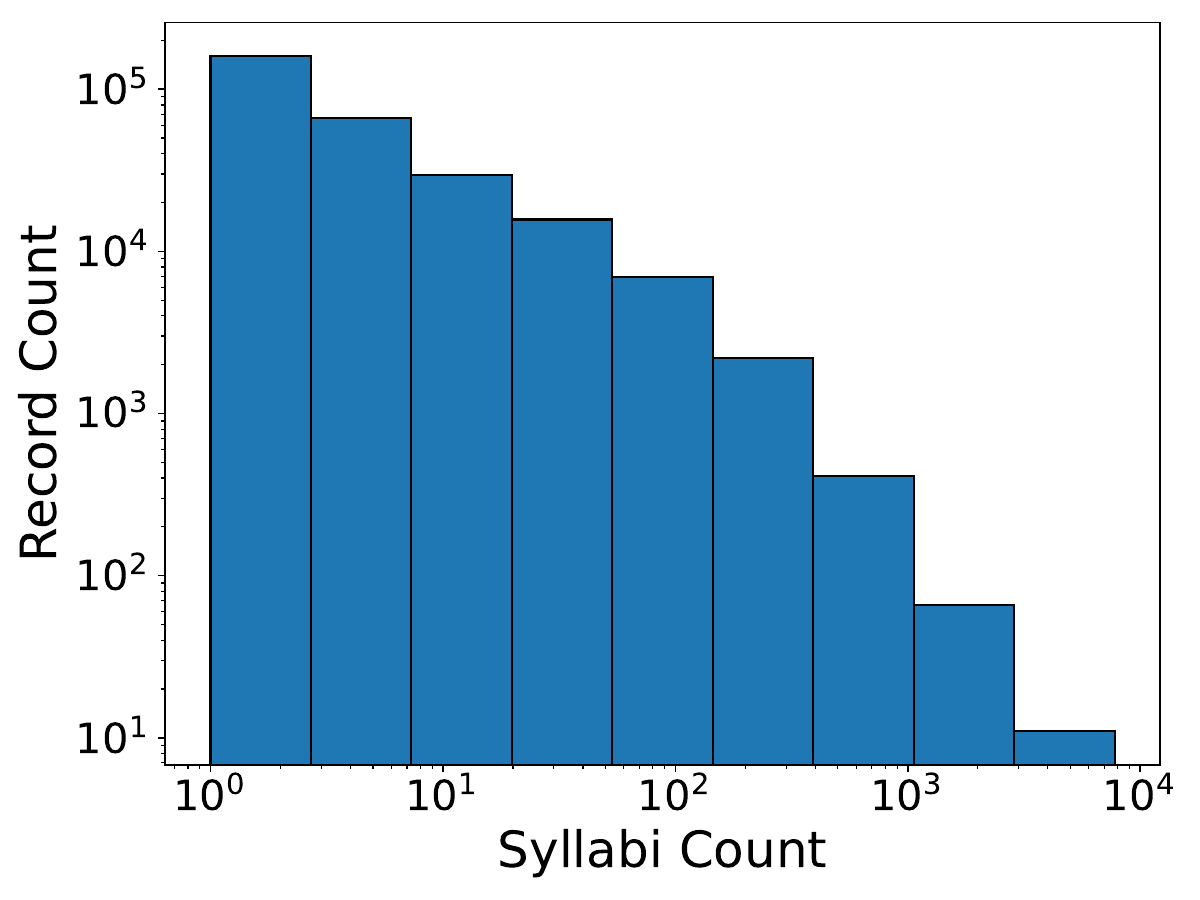}
    \caption{\hl{\textbf{Syllabi count distribution.} The syllabi count distribution of the records (a record is the average score of all syllabi belonging to a FOS, year, and institution triplet).}}
    \label{fig:large_triplet_syllabi_count_dist}
\end{figure}

\begin{figure}[h!]
    \centering
    \includegraphics[trim=0 0 100 0,clip,width=1\textwidth]{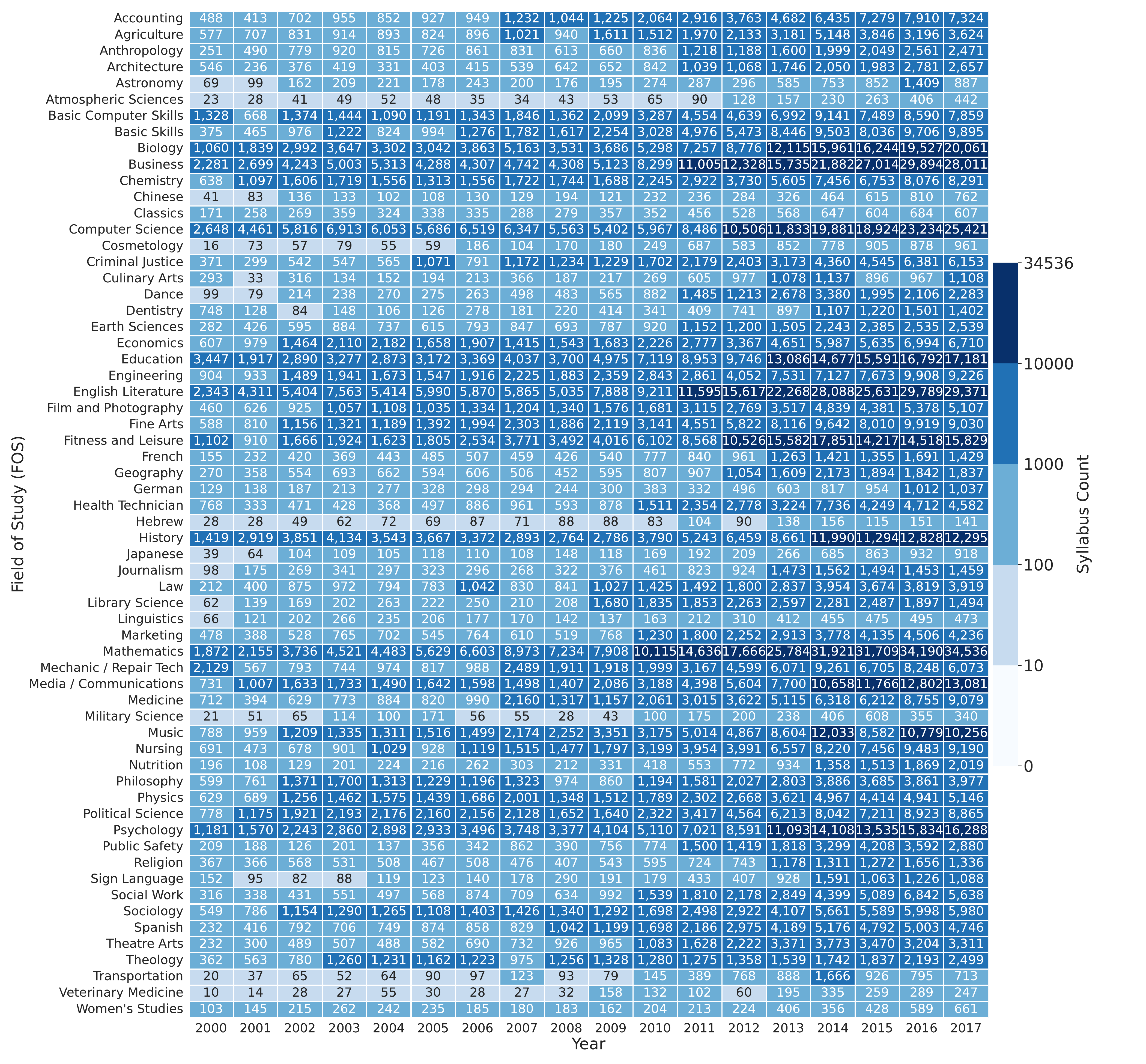}
    \caption{\textbf{Frequency of syllabi per field of study between 2000 and 2017.}}
    \label{fig:fos_per_year_frequency_heatmap}
\end{figure}

\begin{figure}[h!]
    \centering
    \includegraphics[trim=0 0 65 0,clip,width=1\textwidth]{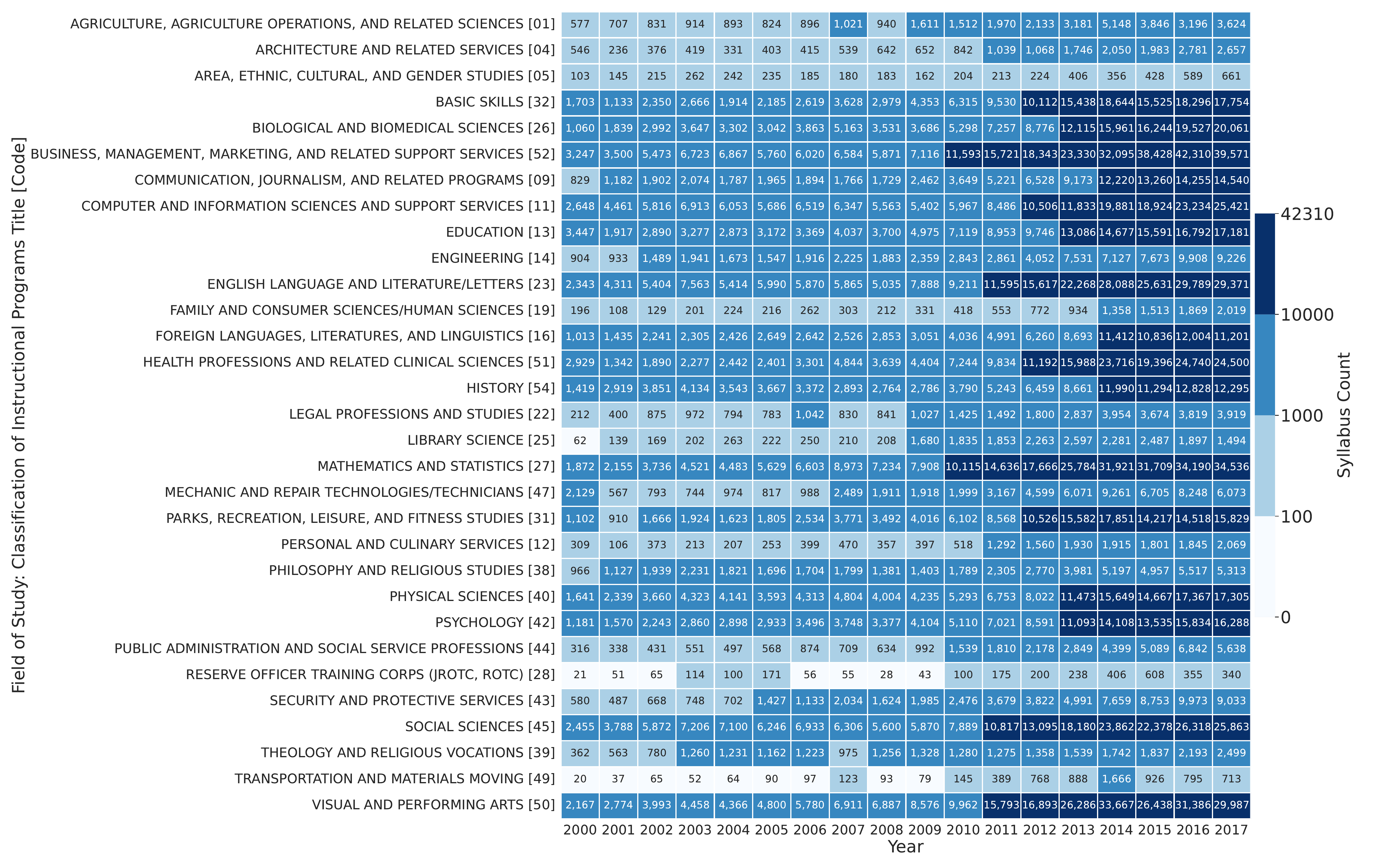}
    \caption{\hl{\textbf{Frequency of syllabi per FOS according to 2-digit CIP 2010 taxonomy  between 2000 and 2017.}}}
    \label{fig:two_digits_fos_yearly_cnt_2000_heatmap}
\end{figure}

Figure{~\ref{fig:large_triplet_syllabi_count_dist}} shows the syllabi count distribution of the aggregated records.
Table{~\ref{tab:fos_frequency}} lists the frequency of syllabi per FOS.
Table~\ref{tab:states_all_stats} details the geographical coverage of the OSP dataset.
Number of educational institutions within each state is obtained from CCIHE.
For example, Texas with $865,973$ syllabi has the largest number of syllabi ($27.85\%$) in the dataset.
According to CCIHE, there are $226$ universities and educational institutions located in Texas, among which $54.42\%$ have at least $8$ syllabi (25\textsuperscript{th} percentile) in the OSP dataset. 
\hl{Moreover, more than $80\%$ of the syllabi belong to the public universities with a majority belonging to the 4-year universities. Nearly $15\%$ are from private not-for-profit, 4-year or above universities
 (see Table{~\ref{tab:sector_stat}} for the frequency and percentage of syllabi per university sector).}
In addition, Table~\ref{tab:institution_frequency} lists the syllabus count per top $100$ universities across all years and all FOS.
Finally, Figure~\ref{fig:fos_per_year_frequency_heatmap} details the number of syllabi per FOS between 2000 and 2017 (see Figure~\ref{fig:two_digits_fos_yearly_cnt_2000_heatmap} for FOS according to 2-digit Classification of Instructional Programs (CIP) 2010 taxonomy).

\begin{table}[h!]
\centering
\small
\resizebox{1.0\textwidth}{!}{
\begin{tabular}{lc|lc|lc}
\hline
\textbf{Field of Study} & \textbf{\# Syllabi} & \textbf{Field of Study} & \textbf{\# Syllabi} & \textbf{Field of Study} & \textbf{\# Syllabi}  \\ \hline
Mathematics             & 258,160             & Accounting              & 51,984              & Religion                & 14,440               \\
English Literature      & 232,065             & Sociology               & 46,836              & French                  & 14,305               \\
Business                & 201,100             & Physics                 & 44,802              & Journalism              & 12,712               \\
Computer Science        & 184,649             & Film and Photography    & 42,690              & Nutrition               & 11,883               \\
Biology                 & 140,187             & Criminal Justice        & 39,805              & Dentistry               & 10,367               \\
Education               & 140,182             & Spanish                 & 39,650              & Culinary Arts           & 9,430                \\
Fitness and Leisure     & 131,262             & Health Technician       & 38,268              & Sign Language           & 8,665                \\
Psychology              & 122,387             & Social Work             & 36,745              & German                  & 8,385                \\
History                 & 107,676             & Philosophy              & 35,583              & Classics                & 7,813                \\
Media / Communications  & 85,561              & Agriculture             & 35,305              & Cosmetology             & 7,291                \\
Music                   & 82,329              & Marketing               & 31,430              & Astronomy               & 7,286                \\
Fine Arts               & 75,722              & Law                     & 31,421              & Transportation          & 7,121                \\
Basic Skills            & 73,362              & Theatre Arts            & 29,087              & Japanese                & 5,456                \\
Engineering             & 70,084              & Theology                & 24,584              & Women's Studies         & 5,237                \\
Political Science       & 69,111              & Public Safety           & 23,931              & Chinese                 & 5,054                \\
Basic Computer Skills   & 68,028              & Earth Sciences          & 21,870              & Linguistics             & 4,859                \\
Nursing                 & 63,603              & Anthropology            & 21,509              & Military Science        & 3,202                \\
Mechanic / Repair Tech  & 62,423              & Library Science         & 20,234              & Atmospheric Sciences    & 2,231                \\
Chemistry               & 61,280              & Dance                   & 19,694              & Veterinary Medicine     & 2,105                \\
Economics               & 56,157              & Architecture            & 19,379              & Hebrew                  & 1,674                \\
Medicine                & 55,161              & Geography               & 17,935              &                         & \multicolumn{1}{l}{} \\ \hline
\end{tabular}
}
\caption{\textbf{Frequency of syllabi per FOS.}}
\label{tab:fos_frequency}
\end{table}

\begin{table}[h!]
\centering
\small
\resizebox{1\textwidth}{!}{
\begin{tabular}{lcccc|lcccc}
\hline
\textbf{Name}        & \textbf{\# Syllabi} & \textbf{\% Syllabi} & \textbf{\# Inst.} & \textbf{\% Covered Inst.} & \textbf{Name}  & \textbf{\# Syllabi} & \textbf{\% Syllabi} & \textbf{\# Inst.} & \textbf{\% Covered Inst.} \\ \hline
Alabama              & 66,548              & 2.14\%              & 60                & 48.33\%                   & Montana        & 6,929               & 0.22\%              & 24                & 37.50\%                   \\
Alaska               & 3,271               & 0.11\%              & 8                 & 50.00\%                   & Nebraska       & 4,609               & 0.15\%              & 34                & 52.94\%                   \\
Arizona              & 16,476              & 0.53\%              & 66                & 31.82\%                   & Nevada         & 16,541              & 0.53\%              & 22                & 31.82\%                   \\
Arkansas             & 9,018               & 0.29\%              & 53                & 56.60\%                   & New Hampshire  & 3,066               & 0.10\%              & 24                & 41.67\%                   \\
California           & 553,589             & 17.80\%             & 425               & 46.82\%                   & New Jersey     & 38,225              & 1.23\%              & 82                & 47.56\%                   \\
Colorado             & 18,718              & 0.60\%              & 62                & 45.16\%                   & New Mexico     & 28,830              & 0.93\%              & 36                & 38.89\%                   \\
Connecticut          & 8,695               & 0.28\%              & 38                & 65.79\%                   & New York       & 77,699              & 2.50\%              & 295               & 43.05\%                   \\
Delaware             & 1,737               & 0.06\%              & 7                 & 57.14\%                   & North Carolina & 76,644              & 2.46\%              & 134               & 41.04\%                   \\
District of Columbia & 13,356              & 0.43\%              & 22                & 36.36\%                   & North Dakota   & 4,120               & 0.13\%              & 20                & 45.00\%                   \\
Florida              & 78,441              & 2.52\%              & 161               & 32.92\%                   & Ohio           & 80,494              & 2.59\%              & 160               & 42.50\%                   \\
Georgia              & 72,955              & 2.35\%              & 107               & 42.06\%                   & Oklahoma       & 35,158              & 1.13\%              & 46                & 69.57\%                   \\
Hawaii               & 19,016              & 0.61\%              & 17                & 52.94\%                   & Oregon         & 24,273              & 0.78\%              & 50                & 56.00\%                   \\
Idaho                & 40,312              & 1.30\%              & 14                & 57.14\%                   & Pennsylvania   & 72,515              & 2.33\%              & 193               & 48.19\%                   \\
Illinois             & 65,455              & 2.10\%              & 152               & 55.26\%                   & Puerto Rico    & 40                  & 0.00\%              & 86                & 2.33\%                    \\
Indiana              & 22,389              & 0.72\%              & 66                & 57.58\%                   & Rhode Island   & 5,322               & 0.17\%              & 15                & 60.00\%                   \\
Iowa                 & 20,861              & 0.67\%              & 56                & 57.14\%                   & South Carolina & 37,259              & 1.20\%              & 66                & 57.58\%                   \\
Kansas               & 14,598              & 0.47\%              & 63                & 50.79\%                   & South Dakota   & 2,880               & 0.09\%              & 21                & 57.14\%                   \\
Kentucky             & 49,872              & 1.60\%              & 59                & 40.68\%                   & Tennessee      & 23,144              & 0.74\%              & 83                & 50.60\%                   \\
Louisiana            & 24,547              & 0.79\%              & 52                & 42.31\%                   & Texas          & 865,973             & 27.85\%             & 226               & 54.42\%                   \\
Maine                & 8,252               & 0.27\%              & 30                & 46.67\%                   & Utah           & 18,612              & 0.60\%              & 23                & 43.48\%                   \\
Maryland             & 157,830             & 5.08\%              & 54                & 70.37\%                   & Vermont        & 5,722               & 0.18\%              & 16                & 62.50\%                   \\
Massachusetts        & 36,102              & 1.16\%              & 106               & 65.09\%                   & Virginia       & 41,186              & 1.32\%              & 108               & 47.22\%                   \\
Michigan             & 120,455             & 3.87\%              & 87                & 63.22\%                   & Washington     & 45,493              & 1.46\%              & 72                & 58.33\%                   \\
Minnesota            & 68,253              & 2.19\%              & 85                & 55.29\%                   & West Virginia  & 5,836               & 0.19\%              & 41                & 34.15\%                   \\
Mississippi          & 7,174               & 0.23\%              & 33                & 75.76\%                   & Wisconsin      & 23,880              & 0.77\%              & 67                & 61.19\%                   \\
Missouri             & 47,594              & 1.53\%              & 93                & 47.31\%                   & Wyoming        & 19,570              & 0.63\%              & 9                 & 66.67\%                   \\ \hline
\end{tabular}
}
\caption{\textbf{Geographical distribution of the OSP dataset.} For each U.S. state ``\# Syllabi'' and ``\% Syllabi'' detail the number and percentage of course syllabi from the universities and institutions located in that state (the sum of ``\% Syllabi'' equals to 100\%).
``\# Inst.'' specifies the number of institutions located in the given state based on the CCIHE.
``\% Covered Inst.'' specifies the percentage of the number of universities with at least 8 course syllabi (25\textsuperscript{th} percentile) in the OSP dataset.}
\label{tab:states_all_stats}
\end{table}

\begin{table}[h!]
\centering
\resizebox{0.8\textwidth}{!}{
\begin{tabular}{lcc}
\hline
\textbf{Sector}                         & \textbf{\# Syllabi} & \textbf{\% Syllabi} \\ \hline
Public, 4-year or above                 & 1,478,045           & 52.60\%             \\
Public, 2-year                          & 813,214             & 28.94\%             \\
Private not-for-profit, 4-year or above & 420,283             & 14.96\%             \\
Not Classified                          & 97,307              & 03.46\%             \\
Private for-profit, 4-year or above     & 1311                & 00.05\%             \\
Private not-for-profit, 2-year          & 31                  & 00.00\%              \\
Private for-profit, 2-year              & 17                  & 00.00\%              \\ \hline
\end{tabular}
}
\caption{\hl{\textbf{Frequency and percentage of syllabi per university sector.}}}
\label{tab:sector_stat}
\end{table}

\begin{table}[h!]
\centering
\small
\resizebox{1\textwidth}{!}{
\begin{tabular}{lc|lc}
\hline
\textbf{Institution Name}                        & \textbf{Count} & \textbf{Institution Name}                       & \textbf{Count} \\ \hline
Alamo Colleges                                   & 160,041        & South Texas College                             & 12,009         \\
University of Maryland University College        & 137,257        & Dallas County Community College District        & 11,641         \\
Amarillo College                                 & 75,198         & Monterey Peninsula College                      & 11,258         \\
Lansing Community College                        & 63,945         & University of Minnesota System                  & 11,220         \\
University of Alabama, Tuscaloosa                & 54,278         & South Plains College                            & 11,184         \\
Texas State Technical College                    & 48,291         & Wilkes University                               & 11,036         \\
Clark State Community College                    & 46,094         & Bellevue College                                & 10,935         \\
Houston Community College System                 & 45,401         & Reedley College                                 & 10,785         \\
Santa Rosa Junior College                        & 35,621         & Modesto Junior College                          & 10,780         \\
Texas A\&M University                            & 33,579         & University of Texas Rio Grande Valley           & 10,653         \\
Rowan-Cabarrus Community College                 & 33,403         & University of Southern California               & 10,480         \\
North Idaho College                              & 31,292         & Kentucky Community and Technical College System & 10,264         \\
University of Texas at Dallas                    & 30,922         & Santa Barbara City College                      & 10,232         \\
University of Georgia                            & 30,349         & Fullerton College                               & 9,993          \\
Texas State University–San Marcos                & 30,051         & Pennsylvania State University                   & 9,303          \\
University of Texas at Arlington                 & 28,651         & Lewis and Clark Community College               & 9,103          \\
San Diego Community College District             & 27,768         & Southwestern Community College                  & 9,004          \\
Western Kentucky University                      & 27,684         & Chaminade University of Honolulu                & 8,866          \\
Park University                                  & 27,096         & Great Basin College                             & 8,837          \\
Sam Houston State University                     & 26,523         & University of Akron                             & 8,787          \\
Stephen F. Austin State University               & 26,170         & University of California, San Diego             & 8,746          \\
University of Michigan–Ann Arbor                 & 25,883         & University of Washington                        & 8,707          \\
Midwestern State University                      & 25,460         & Nova Southeastern University                    & 8,555          \\
Fresno City College                              & 24,786         & University of Colorado Boulder                  & 8,553          \\
George Mason University                          & 24,046         & Rutgers University                              & 8,494          \\
Oral Roberts University                          & 23,950         & University of South Florida                     & 8,193          \\
San Jose State University                        & 23,505         & San Mateo County Community College District     & 8,124          \\
Minnesota State Colleges and Universities System & 23,112         & Palomar College                                 & 8,112          \\
Texas Tech University                            & 22,812         & Westmont College                                & 8,068          \\
McLennan Community College                       & 22,561         & Mt. San Jacinto College                         & 8,015          \\
University of California, Irvine                 & 22,505         & Stony Brook University                          & 7,982          \\
Galveston College                                & 22,077         & New York University                             & 7,959          \\
Tyler Junior College                             & 20,986         & Victoria College                                & 7,919          \\
Clemson University                               & 20,453         & Iowa State University                           & 7,885          \\
University of Texas at Austin                    & 20,387         & University of Maryland, College Park            & 7,842          \\
Collin College                                   & 17,608         & Butte College                                   & 7,728          \\
New Mexico Junior College                        & 16,635         & Merced College                                  & 7,388          \\
Hartnell College                                 & 16,522         & Ventura County Community College District       & 7,368          \\
University of Texas at El Paso                   & 15,842         & Alvin Community College                         & 7,347          \\
Loyola University New Orleans                    & 15,830         & Ohlone College                                  & 7,240          \\
University of North Texas                        & 15,680         & Imperial Valley College                         & 7,174          \\
University of Florida                            & 14,919         & Chaffey College                                 & 7,076          \\
Casper College                                   & 14,665         & Santa Monica College                            & 6,870          \\
University of Texas at San Antonio               & 14,595         & California State University, San Marcos         & 6,865          \\
University of Houston–Clear Lake                 & 14,386         & Palm Beach State College                        & 6,742          \\
Angelo State University                          & 13,663         & Carnegie Mellon University                      & 6,702          \\
Excelsior College                                & 13,517         & El Paso Community College                       & 6,624          \\
Texas A\&M University–Commerce                   & 13,264         & Massachusetts Institute of Technology           & 6,329          \\
Princeton University                             & 13,182         & Florida International University                & 6,287          \\
Santa Ana College                                & 12,969         & Napa Valley College                             & 6,224          \\ \hline
\end{tabular}
}
\caption{\textbf{Most frequent universities.} The syllabus count per top 100 universities across all years and all FOS.}
\label{tab:institution_frequency}
\end{table}

\section*{Technical Validation}
\label{sec:validation}

\subsection*{\hl{Duplicate Analysis}}

\hl{How many duplicate syllabi exist in our dataset?
The syllabi data may have ``duplicates'' because an instructor might teach a course across multiple years with minimal syllabus change, or some introductory courses may have some standard design adopted across institutions. To assess the prevalence of duplicate syllabi in our dataset, we conducted an analysis on the original, disaggregated data. 
Specifically, we compared the DWA skill vectors from syllabi within the same field of study and university across and within various academic years. 
In this context, ``duplicate'' syllabi are defined as those with either identical textual content or learning materials that yield the same similarity score on our NLP framework's assessment.
$25.20\%$ of the total syllabi within a university-major-year are duplicates. This number grows to $31.60\%$ when measuring the total duplicates within a university-major pair.
By taking the differences between these two, we observe that $6.40\%$ of duplicates are across years. 
This relatively low value suggests that instructors are updating their syllabi over time, which leads to differences in skills critical for our analyses. 
These results indicate that majority of the duplicates come from multiple courses in the same major, university and academic year teaching the same content. 
Table{~\ref{tab:duplicate_per_major_df_total_within_across_years}} reports these values by field of study. 
Additionally, Table duplicate\_counts on Figshare{~\cite{ourDataset}}, details the total counts of both original and duplicate syllabi for each university and field of study pairing.}

\begin{table}[h!]
\centering
\small
\resizebox{0.87\textwidth}{!}{
\begin{tabular}{lcc|cc|cc}
\hline
                        & \multicolumn{2}{c|}{\textbf{\begin{tabular}[c]{@{}c@{}}Duplicates per University, Major \\  Combination (Total)\end{tabular}}} & \multicolumn{2}{c|}{\textbf{\begin{tabular}[c]{@{}c@{}}Duplicates per University, Major, \\ Year Combination (Within Year)\end{tabular}}} & \multicolumn{2}{c}{\textbf{\begin{tabular}[c]{@{}c@{}}Duplicates Across Years \\ (Total - Within Year)\end{tabular}}} \\ \hline
\textbf{Field of Study} & \textit{Number}                                          & \textit{Percentage}                                         & \textit{Number}                                                 & \textit{Percentage}                                                 & \textit{Number}                                         & \textit{Percentage}                                         \\ \hline
Library Science         & 12,626                                                   & 64.82\%                                                     & 11,985                                                          & 61.53\%                                                             & 641                                                     & 3.29\%                                                      \\
Transportation          & 3,544                                                    & 56.11\%                                                     & 3,050                                                           & 48.29\%                                                             & 494                                                     & 7.82\%                                                      \\
Public Safety           & 10,987                                                   & 48.91\%                                                     & 9,525                                                           & 42.40\%                                                             & 1,462                                                   & 6.51\%                                                      \\
Mechanic / Repair Tech  & 25,092                                                   & 46.58\%                                                     & 20,504                                                          & 38.06\%                                                             & 4,588                                                   & 8.52\%                                                      \\
Culinary Arts           & 3,665                                                    & 41.95\%                                                     & 2,929                                                           & 33.53\%                                                             & 736                                                     & 8.42\%                                                      \\
Health Technician       & 13,711                                                   & 41.41\%                                                     & 11,019                                                          & 33.28\%                                                             & 2,692                                                   & 8.13\%                                                      \\
Cosmetology             & 2,870                                                    & 40.80\%                                                     & 2,243                                                           & 31.89\%                                                             & 627                                                     & 8.91\%                                                      \\
Basic Computer Skills   & 24,240                                                   & 40.65\%                                                     & 20,633                                                          & 34.61\%                                                             & 3,607                                                   & 6.04\%                                                      \\
Basic Skills            & 25,887                                                   & 39.81\%                                                     & 22,071                                                          & 33.94\%                                                             & 3,816                                                   & 5.87\%                                                      \\
Japanese                & 1,971                                                    & 39.55\%                                                     & 1,601                                                           & 32.12\%                                                             & 370                                                     & 7.43\%                                                      \\
Dentistry               & 3,440                                                    & 39.40\%                                                     & 2,897                                                           & 33.18\%                                                             & 543                                                     & 6.22\%                                                      \\
Criminal Justice        & 14,208                                                   & 38.03\%                                                     & 11,534                                                          & 30.87\%                                                             & 2,674                                                   & 7.16\%                                                      \\
Fitness and Leisure     & 45,192                                                   & 37.65\%                                                     & 34,921                                                          & 29.09\%                                                             & 10,271                                                  & 8.56\%                                                      \\
Music                   & 26,393                                                   & 37.13\%                                                     & 19,150                                                          & 26.94\%                                                             & 7,243                                                   & 10.19\%                                                     \\
Sign Language           & 2,454                                                    & 36.74\%                                                     & 1,920                                                           & 28.75\%                                                             & 534                                                     & 7.99\%                                                      \\
German                  & 2,899                                                    & 36.56\%                                                     & 2,283                                                           & 28.79\%                                                             & 616                                                     & 7.77\%                                                      \\
Atmospheric Sciences    & 760                                                      & 35.70\%                                                     & 566                                                             & 26.59\%                                                             & 194                                                     & 9.11\%                                                      \\
Military Science        & 1,070                                                    & 35.64\%                                                     & 842                                                             & 28.05\%                                                             & 228                                                     & 7.59\%                                                      \\
Mathematics             & 79,952                                                   & 35.40\%                                                     & 64,279                                                          & 28.46\%                                                             & 15,673                                                  & 6.94\%                                                      \\
Spanish                 & 12,352                                                   & 34.92\%                                                     & 10,086                                                          & 28.51\%                                                             & 2,266                                                   & 6.41\%                                                      \\
Nursing                 & 19,529                                                   & 34.52\%                                                     & 16,520                                                          & 29.20\%                                                             & 3,009                                                   & 5.32\%                                                      \\
Dance                   & 5,824                                                    & 34.48\%                                                     & 4,313                                                           & 25.54\%                                                             & 1,511                                                   & 8.94\%                                                      \\
English Literature      & 71,645                                                   & 34.41\%                                                     & 63,061                                                          & 30.28\%                                                             & 8,584                                                   & 4.13\%                                                      \\
Nutrition               & 3,590                                                    & 33.39\%                                                     & 2,868                                                           & 26.68\%                                                             & 722                                                     & 6.71\%                                                      \\
Accounting              & 15,836                                                   & 32.59\%                                                     & 12,447                                                          & 25.61\%                                                             & 3,389                                                   & 6.98\%                                                      \\
Business                & 61,178                                                   & 32.22\%                                                     & 49,171                                                          & 25.90\%                                                             & 12,007                                                  & 6.32\%                                                      \\
Computer Science        & 53,616                                                   & 32.20\%                                                     & 42,614                                                          & 25.59\%                                                             & 11,002                                                  & 6.61\%                                                      \\
Media / Communications  & 24,793                                                   & 31.92\%                                                     & 20,681                                                          & 26.63\%                                                             & 4,112                                                   & 5.29\%                                                      \\
Agriculture             & 9,459                                                    & 31.74\%                                                     & 6,387                                                           & 21.43\%                                                             & 3,072                                                   & 10.31\%                                                     \\
Astronomy               & 1,841                                                    & 31.49\%                                                     & 1,374                                                           & 23.50\%                                                             & 467                                                     & 7.99\%                                                      \\
Biology                 & 39,526                                                   & 31.36\%                                                     & 30,387                                                          & 24.11\%                                                             & 9,139                                                   & 7.25\%                                                      \\
Fine Arts               & 21,049                                                   & 30.81\%                                                     & 16,321                                                          & 23.89\%                                                             & 4,728                                                   & 6.92\%                                                      \\
Medicine                & 15,591                                                   & 30.63\%                                                     & 12,420                                                          & 24.40\%                                                             & 3,171                                                   & 6.23\%                                                      \\
French                  & 3,931                                                    & 29.97\%                                                     & 2,962                                                           & 22.58\%                                                             & 969                                                     & 7.39\%                                                      \\
Earth Sciences          & 5,946                                                    & 29.74\%                                                     & 4,226                                                           & 21.14\%                                                             & 1,720                                                   & 8.60\%                                                      \\
Chemistry               & 16,349                                                   & 29.72\%                                                     & 11,917                                                          & 21.67\%                                                             & 4,432                                                   & 8.05\%                                                      \\
Physics                 & 11,255                                                   & 29.14\%                                                     & 8,148                                                           & 21.10\%                                                             & 3,107                                                   & 8.04\%                                                      \\
Theatre Arts            & 7,853                                                    & 28.94\%                                                     & 6,142                                                           & 22.63\%                                                             & 1,711                                                   & 6.31\%                                                      \\
Chinese                 & 1,337                                                    & 28.83\%                                                     & 1,003                                                           & 21.63\%                                                             & 334                                                     & 7.20\%                                                      \\
Film and Photography    & 11,182                                                   & 28.48\%                                                     & 8,784                                                           & 22.37\%                                                             & 2,398                                                   & 6.11\%                                                      \\
History                 & 26,505                                                   & 27.77\%                                                     & 21,129                                                          & 22.14\%                                                             & 5,376                                                   & 5.63\%                                                      \\
Law                     & 7,545                                                    & 27.41\%                                                     & 5,542                                                           & 20.14\%                                                             & 2,003                                                   & 7.27\%                                                      \\
Veterinary Medicine     & 370                                                      & 26.93\%                                                     & 270                                                             & 19.65\%                                                             & 100                                                     & 7.28\%                                                      \\
Engineering             & 17,067                                                   & 26.91\%                                                     & 12,318                                                          & 19.42\%                                                             & 4,749                                                   & 7.49\%                                                      \\
Marketing               & 7,965                                                    & 26.64\%                                                     & 6,150                                                           & 20.57\%                                                             & 1,815                                                   & 6.07\%                                                      \\
Sociology               & 11,411                                                   & 26.57\%                                                     & 9,199                                                           & 21.42\%                                                             & 2,212                                                   & 5.15\%                                                      \\
Religion                & 3,509                                                    & 25.03\%                                                     & 2,792                                                           & 19.91\%                                                             & 717                                                     & 5.12\%                                                      \\
Psychology              & 28,009                                                   & 24.83\%                                                     & 21,883                                                          & 19.40\%                                                             & 6,126                                                   & 5.43\%                                                      \\
Classics                & 1,717                                                    & 24.72\%                                                     & 1,110                                                           & 15.98\%                                                             & 607                                                     & 8.74\%                                                      \\
Economics               & 12,356                                                   & 24.14\%                                                     & 8,904                                                           & 17.40\%                                                             & 3,452                                                   & 6.74\%                                                      \\
Social Work             & 8,340                                                    & 23.01\%                                                     & 6,806                                                           & 18.78\%                                                             & 1,534                                                   & 4.23\%                                                      \\
Geography               & 3,681                                                    & 22.64\%                                                     & 2,550                                                           & 15.68\%                                                             & 1,131                                                   & 6.96\%                                                      \\
Philosophy              & 6,881                                                    & 22.04\%                                                     & 5,391                                                           & 17.27\%                                                             & 1,490                                                   & 4.77\%                                                      \\
Political Science       & 13,088                                                   & 21.37\%                                                     & 10,097                                                          & 16.49\%                                                             & 2,991                                                   & 4.88\%                                                      \\
Hebrew                  & 334                                                      & 20.99\%                                                     & 185                                                             & 11.63\%                                                             & 149                                                     & 9.36\%                                                      \\
Journalism              & 2,552                                                    & 20.51\%                                                     & 1,871                                                           & 15.04\%                                                             & 681                                                     & 5.47\%                                                      \\
Anthropology            & 3,993                                                    & 19.94\%                                                     & 3,175                                                           & 15.86\%                                                             & 818                                                     & 4.08\%                                                      \\
Education               & 26,628                                                   & 19.70\%                                                     & 21,342                                                          & 15.79\%                                                             & 5,286                                                   & 3.91\%                                                      \\
Architecture            & 3,115                                                    & 17.76\%                                                     & 2,303                                                           & 13.13\%                                                             & 812                                                     & 4.63\%                                                      \\
Women's Studies         & 872                                                      & 17.38\%                                                     & 726                                                             & 14.47\%                                                             & 146                                                     & 2.91\%                                                      \\
Theology                & 3,511                                                    & 14.75\%                                                     & 2,098                                                           & 8.82\%                                                              & 1,413                                                   & 5.93\%                                                      \\
Linguistics             & 597                                                      & 12.99\%                                                     & 374                                                             & 8.14\%                                                              & 223                                                     & 4.85\%       \\\hline                                              
\end{tabular}
}
\caption{\hl{\textbf{Duplicate syllabi per field of study.} A duplicate refers to a set of DWA vector with the same scores. The first column shows the total number and percentage of the duplicates obtained from each institution-FOS combinations. 
The second column shows the number and percentage of the duplicates within a year calculated based on institution-FOS-year combinations. 
And the last column is the total duplicates minus the the within year duplicates. 
The duplicate column shows the average of duplicate syllabi (i.e., having the same DWA skill vector) within the same field of study and university across different years.}}
\label{tab:duplicate_per_major_df_total_within_across_years}

\end{table}

\subsection*{\hl{Representativeness Analysis}}

\hl{How representative is our data of US higher education? 
A recent study using the OSP dataset{~\cite{biasi2022eduInov}} shows that the syllabi sample consistently represents about $5\%$ of all courses taught in US institutions from 1998 to 2018. In their analysis of courses from $161$ representative US institutions, they find that the sample slightly overrepresents Ivy-Plus schools but is broadly representative in terms of field and course level distribution. Despite this bias, the dataset adequately reflects U.S. higher education offerings during the period, with no significant bias in financial resources or student demographics.}

\begin{figure}[h!]
    \centering
    \includegraphics[trim=0 0 0 0,clip,width=1\textwidth]{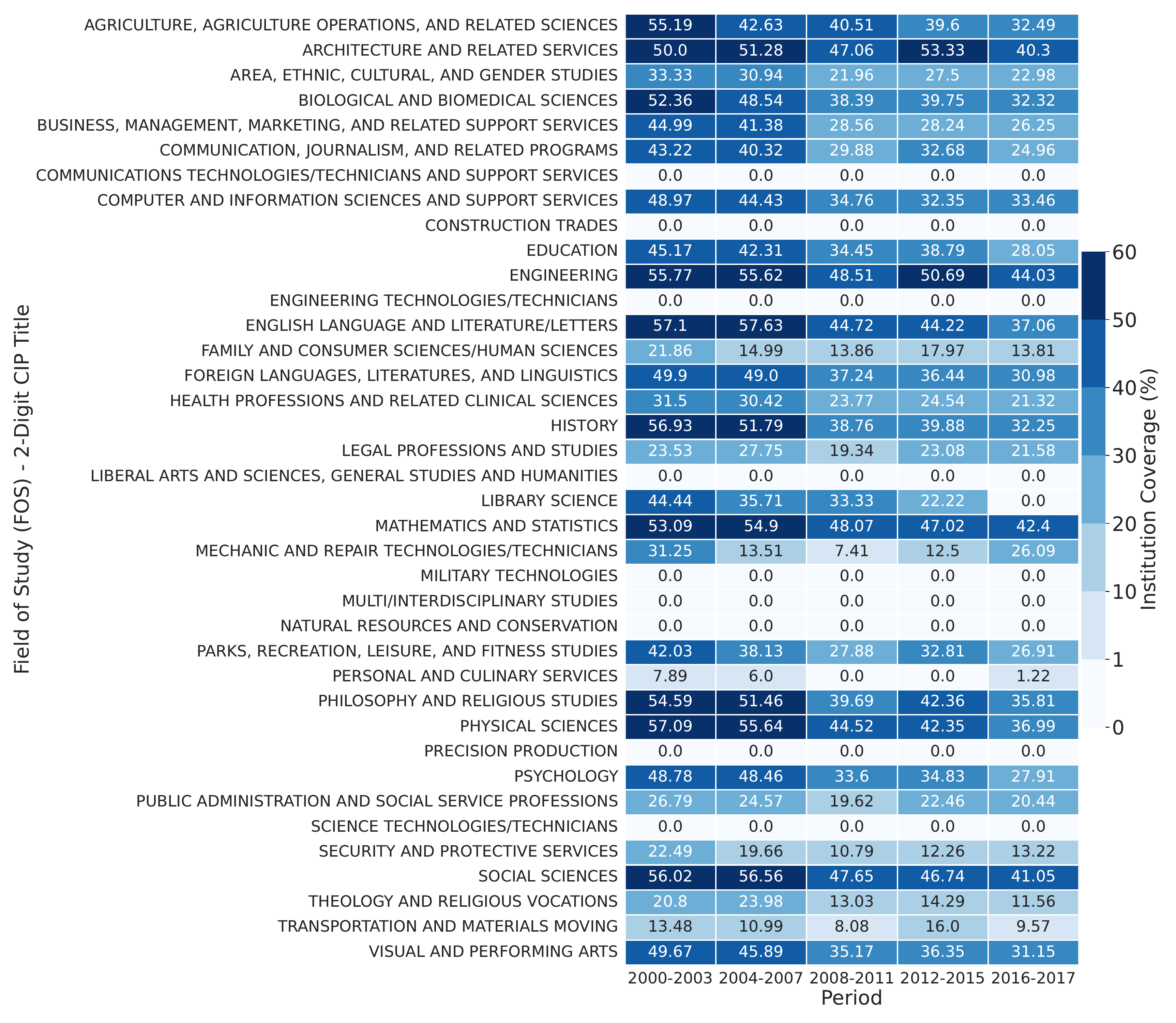}
    \caption{\hl{\textbf{Trends in the coverage of institutions across different fields of study.}
    The heatmap visualizes the coverage trends for various fields of study (2-digit CIP titles) across six time periods (2000-2003, 2004-2007, 2008-2011, 2012-2015, and 2016-2017). 
    The coverage is the percentage of the number of institutions with at least one syllabus from a institution-FOS-year combination divided by the total number of institutions with a positive number of bachelor's degrees from the same institution-FOS-year combination obtained from IPEDS.}}
    \label{fig:representativeness_syllabi_grads}
\end{figure}

\hl{Moreover, we calculate the coverage of institutions across different FOS according to the number of Bachelor's degrees awarded in a FOS across US institutions between 2000 and 2017 obtained from IPEDS completions data (}\url{https://nces.ed.gov/ipeds/datacenter/DataFiles.aspx?year=-1&surveyNumber=3&sid=6758f146-5ae5-44a0-8982-9821e70a8757&rtid=7}). 
\hl{We calculate the coverage as the percentage of the number of institutions with at least one syllabus from a institution-FOS-year combination divided by the total number of institutions with a positive number of bachelor's degrees from the same institution-FOS-year combination obtained from IPEDS. 
Fields such as ``Engineering'' and ``Social Sciences'' show a consistent coverage over $40\%$. 
On the other hand, several fields like ``Military Technologies'' and ``Natural Resources and Conservation'' have consistently zero coverage (see Fig.{~\ref{fig:representativeness_syllabi_grads}}). 
Table ipeds\_2digit\_grad\_2000\_2017 on Figshare {\cite{ourDataset}} provides the number of graduates per university and 2-digit CIP code between 2000 and 2017 downloaded from IPEDS.}

\subsection*{\hl{Sufficient Number of Syllabi}}

\begin{figure}[h!]
    \centering
    \includegraphics[trim=0 175 0 0,clip,width=0.95\textwidth]{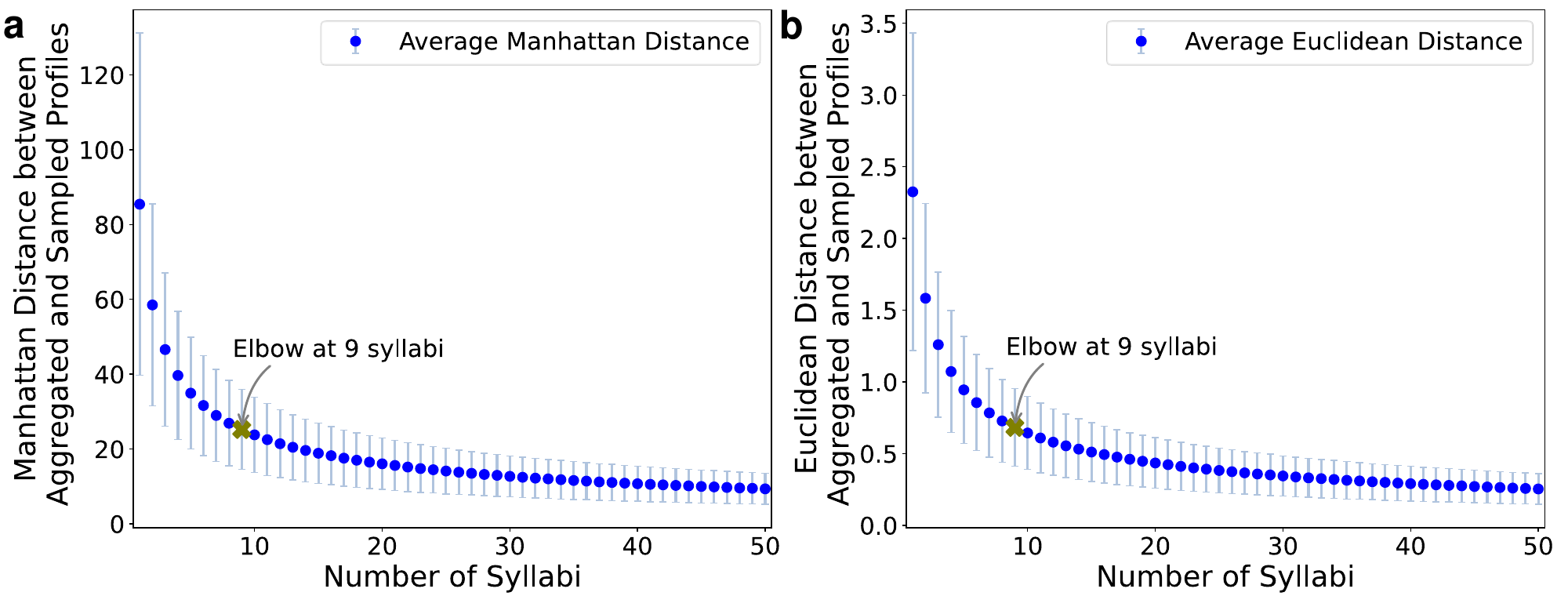}
    \caption{\hl{\textbf{Identification of the sufficient number of syllabi.} 
    The average (a) Manhattan Distance and (b) Euclidean Distance between the aggregated and a given number of randomly selected syllabi. 
    Each point represents the mean distance for a given number of syllabi, with their corresponding error bar.
    The elbow points, marked with an olive `X' and annotated, indicate the sufficient number of syllabi (here, 9 syllabi) where the rate of decrease in distance significantly slows down. 
    These points help identify the threshold beyond which additional syllabi contribute minimally to reducing the distance, providing a practical cutoff for data aggregation. Note that the x-axis is limited to 50 for the visualization purpose.}}
    \label{fig:n_syllabi_elbow_with_label}
\end{figure}

\begin{figure}[h!]
    \centering
    \includegraphics[trim=0 0 0 0,clip,width=0.97\textwidth]{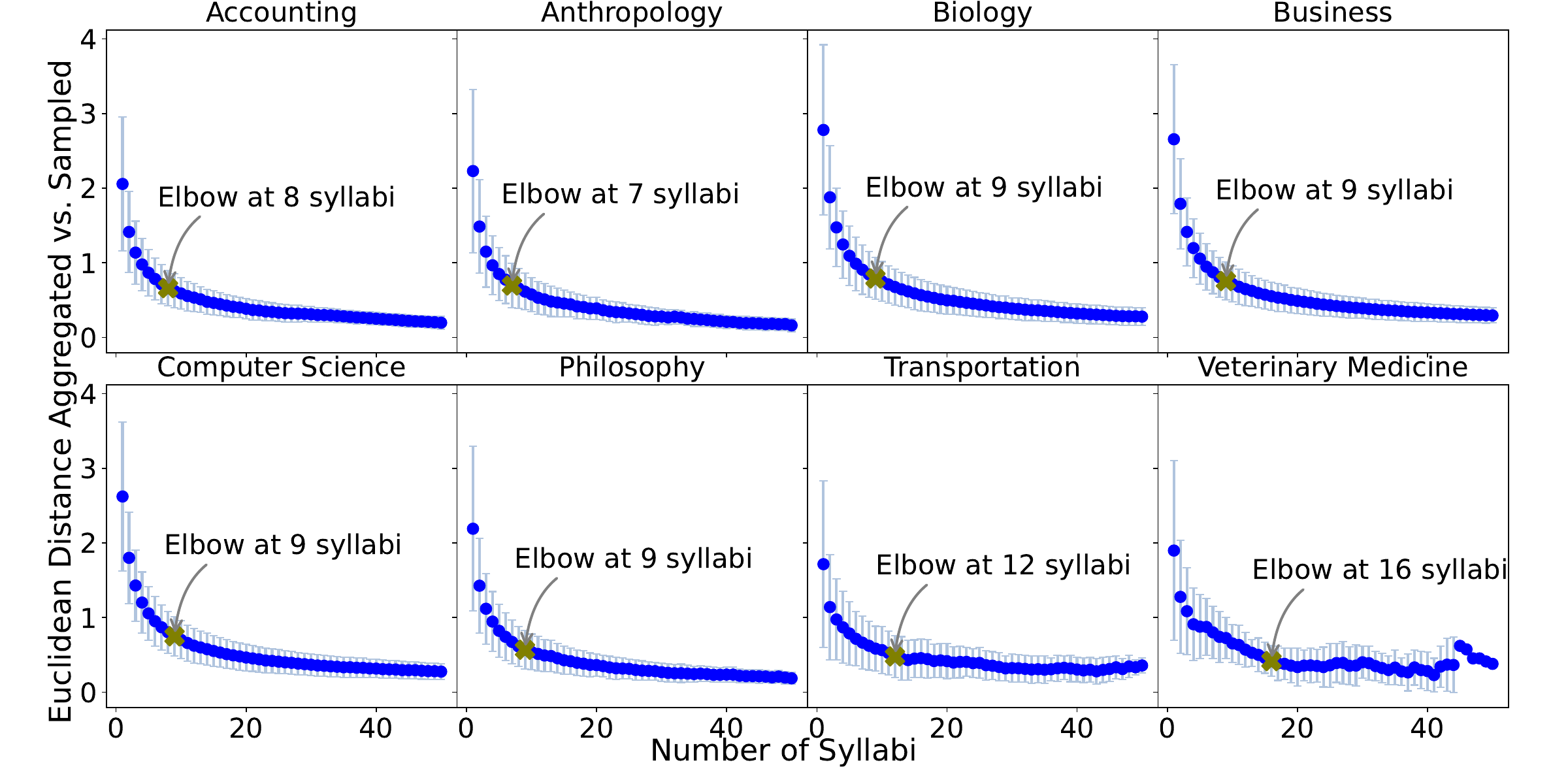}
    \caption{\hl{\textbf{Identification of the sufficient number of syllabi for sample of FOS.} 
    The average Euclidean Distance between the the aggregated and a given number of randomly selected syllabi within each FOS. 
    Each point represents the mean distance for a given number of syllabi, with their corresponding error bar. 
    The elbow points, marked with an olive `X' and annotated, indicate the sufficient number of syllabi (e.g., 8 syllabi for Accounting and 10 syllabi for Architecture) where the rate of decrease in distance significantly slows down. These points help identify the threshold beyond which additional syllabi contribute minimally to reducing the distance, providing a practical cutoff for data aggregation. Note that the x-axis is limited to 50 for the visualization purpose.}}
    \label{fig:n_syllabi_elbow_per_major_sample}
\end{figure}

\hl{How stable are our estimates of the skill content associated with each FOS and institution in each year? 
We analyze DWA institution-FOS-year combinations with at least two syllabi post-2000 (i.e., $2,686,066$ syllabi). 
For each institution-FOS-year combination, we calculate the Manhattan and Euclidean distances between the aggregated published skill profile (i.e., the 
average scores of the complete set of syllabi) and the skill profile of a randomly selected subsample of syllabi (from one syllabus up to the maximum number of syllabi for the given institution-FOS-year combination minus one).
We perform ten trials for each subset size.
Finally, we average the distances of each subset size for all syllabi and calculate their distances.
Figure{~\ref{fig:n_syllabi_elbow_with_label}} illustrates this analysis using Manhattan (Fig.{~\ref{fig:n_syllabi_elbow_with_label}}a) and Euclidean (Fig.{~\ref{fig:n_syllabi_elbow_with_label}}b) distances between the aggregated and sampled profiles.
Each point in the plots represents the mean distance for a given number of syllabi and the error bars represent a $95\%$ confidence interval.
The elbow points{~\cite{elbowNg, elbowReview}}, annotated with ``X'', indicate where the rate of decrease in distance significantly slows down thus identifying the sufficient number of syllabi is equal to $9$. 
Out of $281,153$ published institution-FOS-year combinations, $49,750$ have at least $9$ syllabi ($17.69\%$). 
Note that the minimum number of syllabi (i.e., $9$) is obtained using nearly all the available syllabi and the y-axes of the figures are limited to $50$ for the visualization purpose.
Moreover, to see how the sufficient number might vary within each FOS, we redo the mentioned procedure for all the institution-FOS-year combinations within a given FOS.
The minimum number of syllabi for a majority of the fields of study is between $8$ and $10$ with some exceptions for Transportation and Veterinary Medicine (see Fig.{~\ref{fig:n_syllabi_elbow_per_major_sample}} for examples and Figure euclidean\_n\_syllabi\_elbow\_per\_major on Figshare{~\cite{ourDataset}} for the complete list.).
We publish all the calculated distances in Table manhattan\_euclidean\_distances on Figshare{~\cite{ourDataset}}.
}

\begin{figure}[h!]
    \centering
    \includegraphics[trim=0 10 0 0,clip,width=0.60\textwidth]{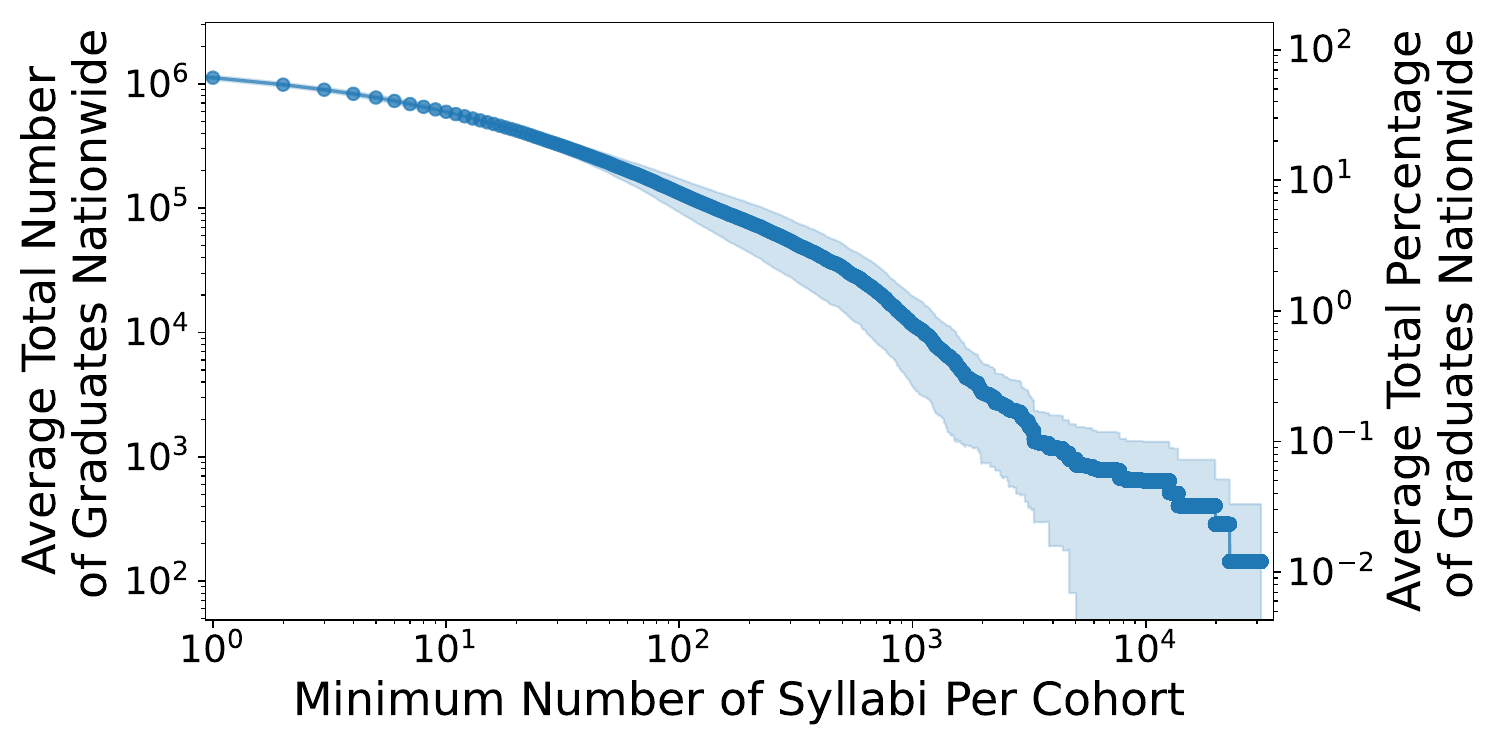}
    \caption{\hl{\textbf{The relationship between the minimum number of syllabi per cohort and the average total number of graduates nationwide between 2000 and 2017.}
    The x-axis shows the minimum number of syllabi per cohort (institution, FOS, period) and the y-axes show the average total number (left) and percentage (right) of graduates nationwide.    
    The plot shows how the total number of graduates decreases as the minimum number of syllabi per cohort increases. The shaded area corresponds to the 95\% confidence interval.}}
    \label{fig:min_syll_cnt_tot_grads}
\end{figure}

\hl{Lastly, how does the minimum number of syllabi per cohort relate to the total number of graduates with available syllabi in our dataset?
We explore by counting the number of graduates per institution-FOS combination between 2003 and 2017 while varying the minimum number of syllabi from different cohorts between 2000 and 2017 (see Fig.{~\ref{fig:min_syll_cnt_tot_grads}}).
The first cohort (2003) contains the syllabi between 2000 to 2003.
For example, considering all the institution-FOS-year combinations with at least one syllabus allows us to analyze the course materials of nearly $1.1$ million nationwide Bachelor's degree graduates per year.
However, restricting to institution-FOS-year combinations with at least $9$ syllabi narrows the analysis to around $622,000$ graduates, i.e., around $35.46\%$ of graduates per year.}

\subsection*{Qualitative Analysis of the Inferred Workplace Activities}

\begin{figure}[h!]
    \centering
    \includegraphics[trim=40 0 60 0,clip,width=0.90\textwidth]{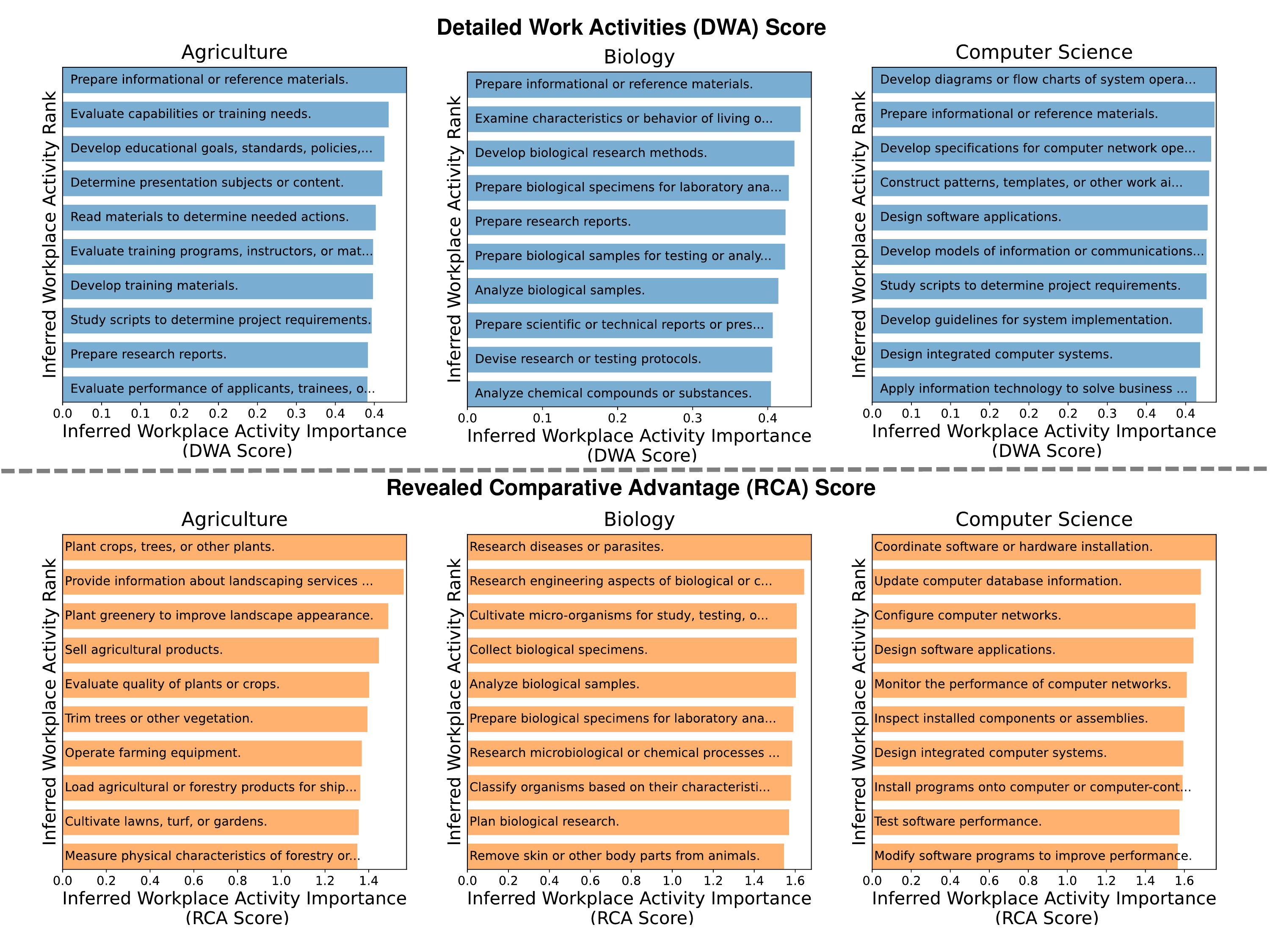}
    \caption{\textbf{The DWAs most strongly associated with Agriculture, Biology, and Computer Science.} (\textit{top}) Top $10$ inferred workplace activities with the highest DWA scores.
    (\textit{bottom}) Top DWAs according to their RCA scores.}
    \label{fig:dwa_mean_rca_3_samples}
\end{figure}

As a face-validity check of \texttt{Syllabus2O\textsuperscript{*}NET}, we list the ten DWAs that are most strongly associated with three example fields of study (FOS): Agriculture, Biology, and Computer Science (see Fig.~\ref{fig:dwa_mean_rca_3_samples}).
Some DWAs (e.g., ``Prepare informational or reference materials'') are common across many FOS and, therefore, obscure the DWAs that most distinguish one FOS from the others.
To normalize for ubiquitous DWAs, we also present the DWAs with the greatest revealed comparative advantage (RCA) in each field (see Section ``Skill Normalization'').
For instance, ``Plant crops, trees, or other plants'' emerges as the foremost skill in \textit{Agriculture}, ``Research diseases or parasites'' is predominant in \textit{Biology}, and ``Coordinate software or hardware installation'' is leading in \textit{Computer Science} according to RCA scores.
Top 10 DWA per FOS contains similar results for each FOS \hl{(see Table top10\_DWA\_per\_FOS on Figshare}~\cite{ourDataset}).

\subsection*{Relating Fields of Study from Skill Similarity}

\begin{figure}[h!]
    \centering
    \includegraphics[trim=15 110 205 130,clip,width=0.80\textwidth]{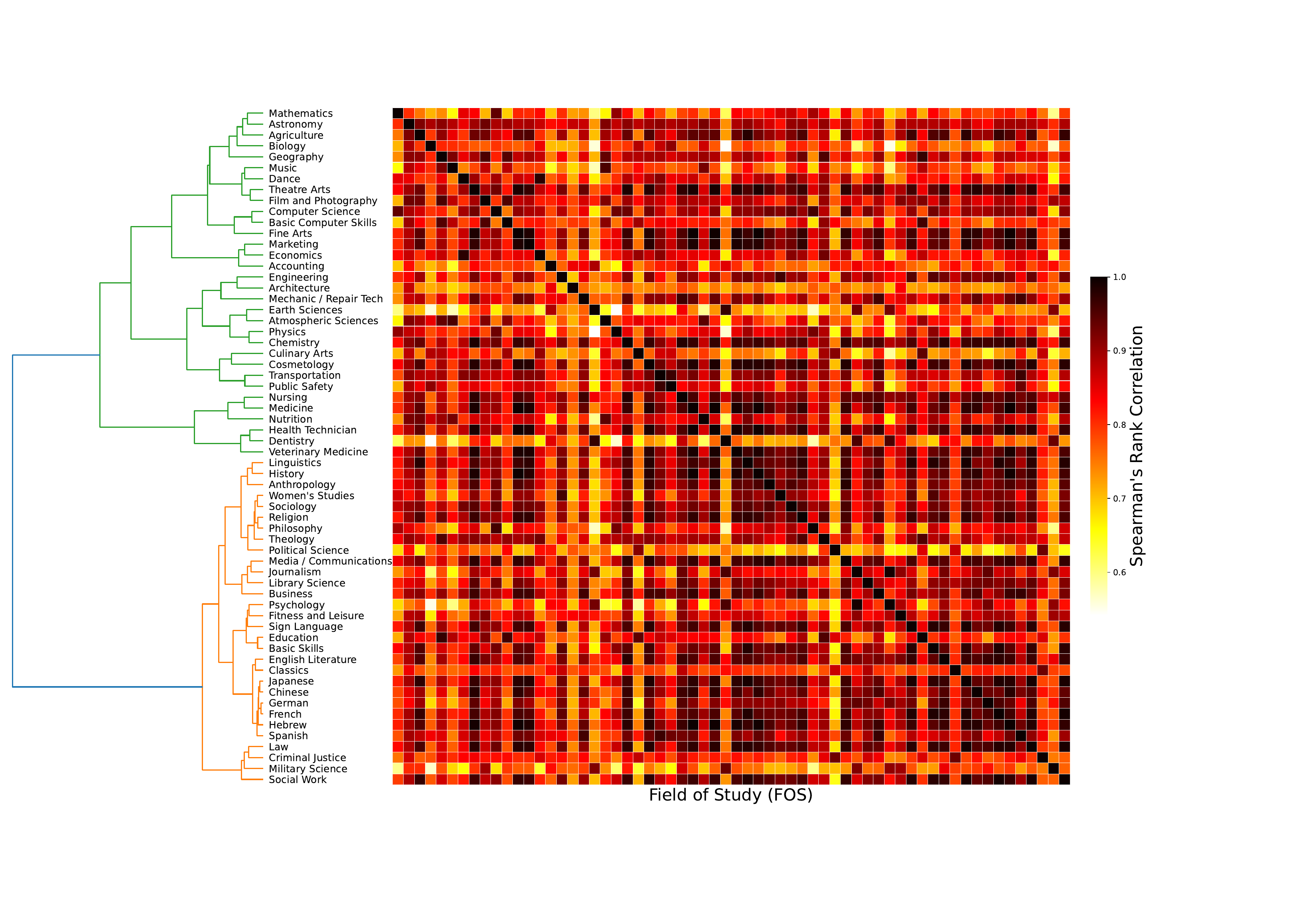}
    \caption{\textbf{The similarity of FOS according to their DWA profiles.} The heatmap shows the Spearman’s rank correlation between the fields of study and the dendrogram represents the results of hierarchically clustering similar FOS. \hl{The dendrogram on the left side organizes the FOS into clusters based on their similarity. Each branch point (node) indicates a point where two branches merge, showing the hierarchical relationship between the fields of study. Fields of study that cluster together (i.e., merge at lower levels) are more similar to each other in terms of their task profiles. For instance, closely related fields like ``Physics'' and ``Chemistry'' are grouped together.}}
    \label{fig:DWA_field_name_dendogram}
\end{figure}

\begin{figure}[h!]
    \centering
    \includegraphics[trim=15 110 200 130,clip,width=0.80\textwidth]{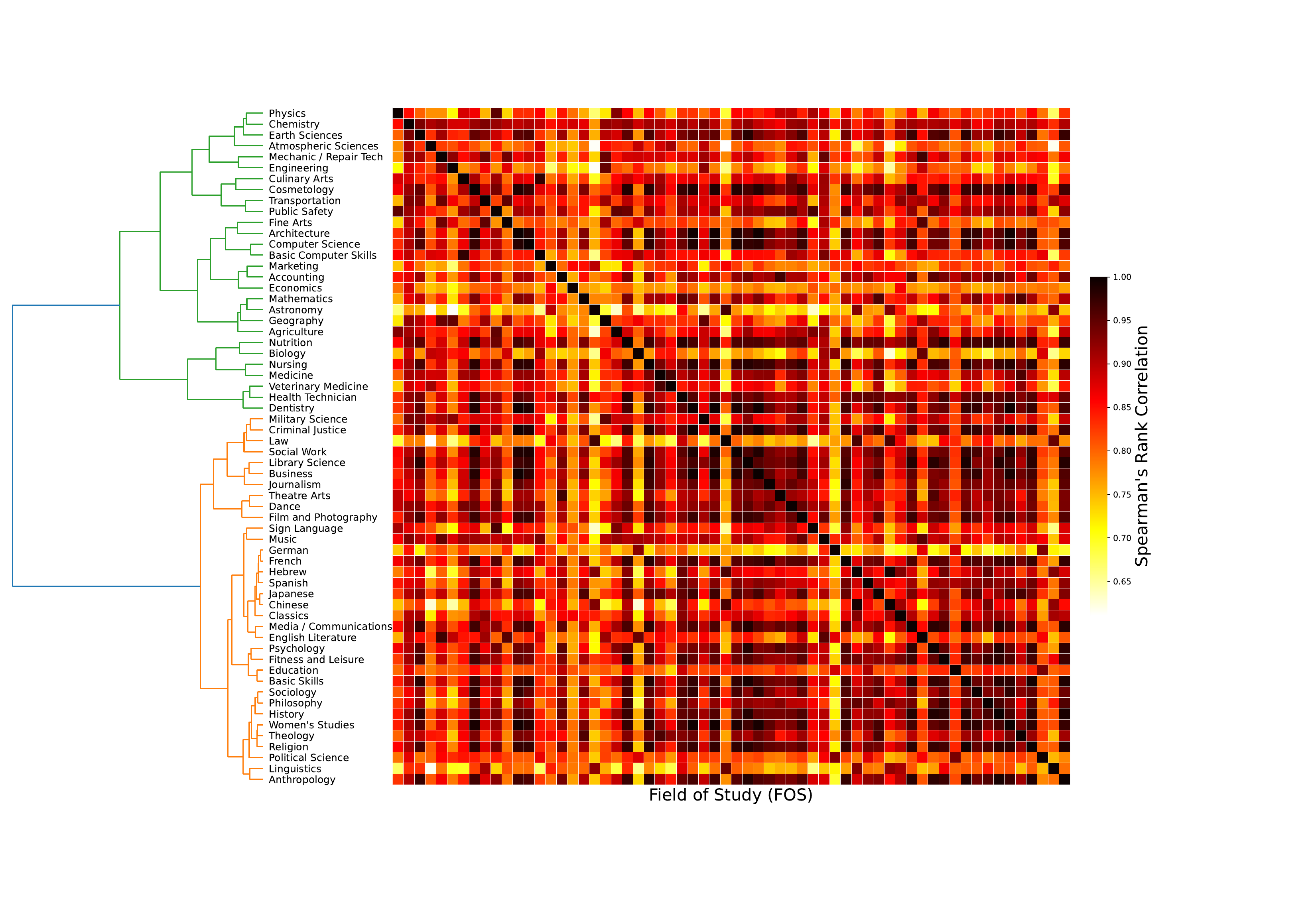}
    \caption{\textbf{The similarity of FOS according to their Task profiles.} The heatmap shows the Spearman’s rank correlation between the fields of study and the dendrogram represents the results of hierarchically clustering similar FOS. \hl{The dendrogram on the left side organizes the FOS into clusters based on their similarity. Each branch point (node) indicates a point where two branches merge, showing the hierarchical relationship between the fields of study. Fields of study that cluster together (i.e., merge at lower levels) are more similar to each other in terms of their task profiles. For instance, closely related fields like ``Physics'' and ``Chemistry'' are grouped together.}}
    \label{fig:Task_field_name_dendogram}
\end{figure}

How similar are fields of study based on their skills?
Following existing work~\cite{chau2023connecting}, \hl{we employ agglomerative hierarchical clustering technique}~\cite{johnson1967hierarchical} \hl{on the DWA vector representations of academic majors, aiming to elucidate their hierarchical relationships. 
Hierarchical clustering generates a nested sequence of clusters, allowing for an in-depth exploration of clusters at varying levels of granularity without predefining a specific number of categories (in this context, groups of majors). 
In this framework, FOS are deemed similar if they share the same work activities} (see Figure~\ref{fig:DWA_field_name_dendogram} for DWAs \hl{and Figure}~\ref{fig:Task_field_name_dendogram} \hl{for Tasks).}

The resulting dendrogram offers another face-validity check as similar FOS (e.g., STEM majors) tend to require similar DWAs. 
For instance, Marketing and Economics are closely related, as are Linguistics and History. 
Notably, just before the final clustering step, which amalgamates all majors (indicated in blue), two predominant clusters are discernible: one representing technical majors including those in STEM (in green) and the other humanities-based majors (in orange). 
For example, although Film and Photography is not a STEM-designated program, it requires skills that are common in STEM fields, such as ``Draw detailed or technical illustrations'' and ``Design video game features or details'' \hl{(see Table top10\_DWA\_per\_FOS on Figshare}~\cite{ourDataset}).

\subsection*{Dynamic Differences Between Inferred Workplace Activities and Labor Market Workplace Activities}

How responsive are the skills taught in higher education to the skills required in the U.S. labor market?
A ``skill mismatch'' may occur if higher education fails to adapt to the demands of the labor market~\cite{aloysius2018enhancing,pujol2015competences,jackson2022relative} (e.g., by teaching more theoretical skills than practical skills~\cite{aloysius2018enhancing}).
Our dataset naturally offers an avenue of examining this mismatch as the scores are computed via comparison between taught content in higher education and skills in the labor market defined by the federal government.
To validate this utility, we perform a similar analysis to that done by B\"{o}rner and colleagues~\cite{borner2018skill}. 
In their study, skill mentions were identified in course syllabi and in job postings to compare skills taught and demanded in computer science related fields in the US, where skills came from a skill taxonomy established by the Burning Glass Technologies. 
Then, Kullback-Leibler (KL) divergence was used to quantify the difference between skill distributions in course syllabi and those in job postings, and between different time periods.

\begin{figure}[h!]
    \centering
    \includegraphics[trim=0 95 0 0,clip,width=1\textwidth]{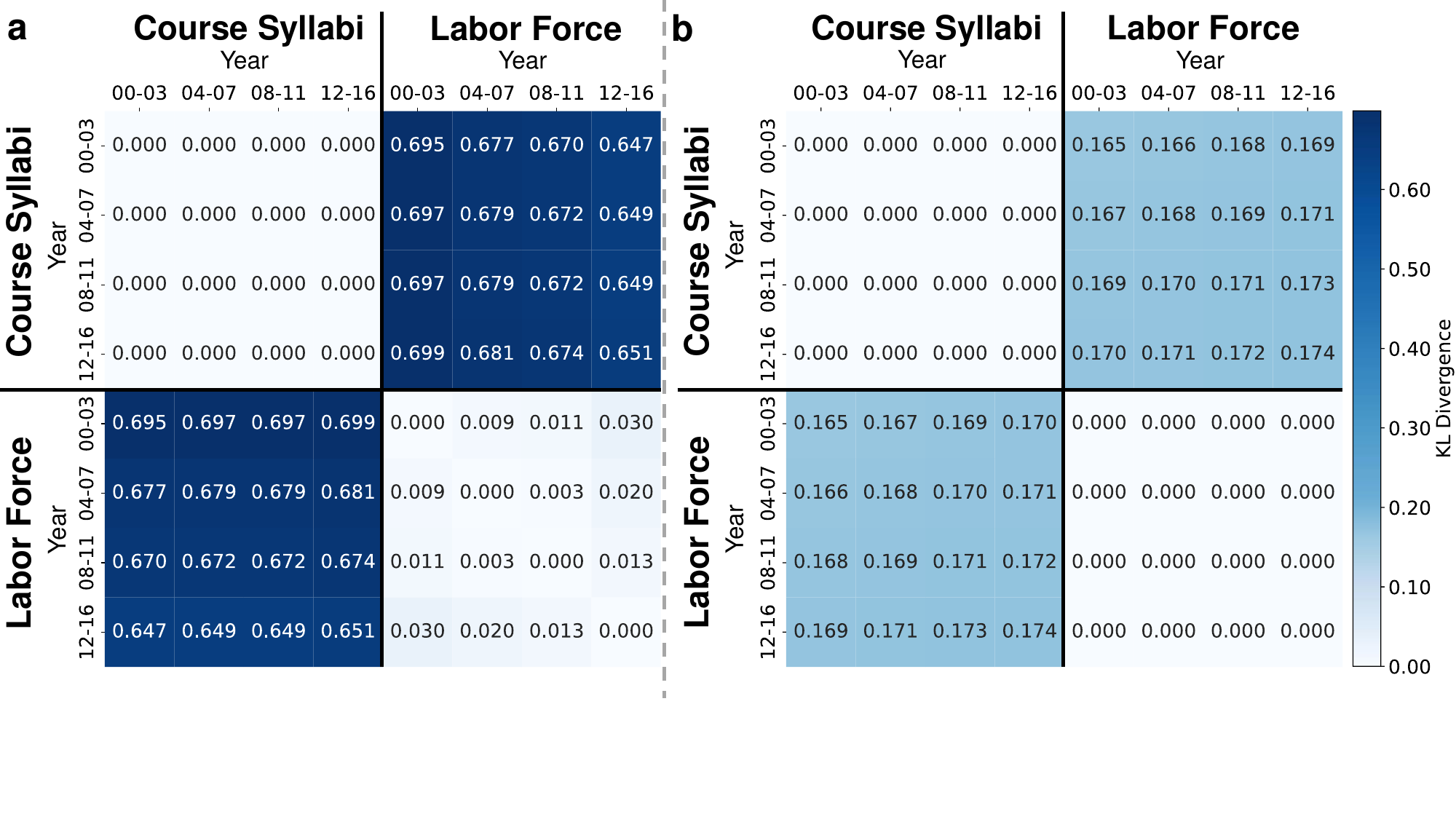}
    \caption{\textbf{Dynamic differences between skill (DWA) distributions in course syllabi and the labor force.}  The matrix reports pairwise Kullback-Leibler (KL) divergence between course syllabi and the labor force for (a) syllabi from all FOS and employment-weighted O*NET DWA profiles for occupations requiring a university degree, and (b) Computer Science and Mathematics course syllabi and employment-weighted O*NET DWA profiles for Computer and Mathematical Occupations (SOC 15-0000). 
    \hl{
    The off-diagonal elements in the Course Syllabi and Labor Force cells are looking at the correlation between the DWAs in those time periods and the syllabi in those time periods. These are not required to be symmetric, meaning that $D_{KL}(P \parallel Q)$ is not necessarily equal to $D_{KL}(Q \parallel P)$.}
}
    \label{fig:klDivergence_all_vs_csmath}
\end{figure}

With the Course-Skill Atlas dataset, we have measures of skills taught in course syllabi. On the labor market side, each occupation in the economy can be characterized by the same measure of skill, in this case, DWAs. 
To generate a measure of the economy-wide skill demand, 
\hl{we weight by the national employment share associated with each occupation from the US Bureau of Labor Statistics (BLS) Occupation Employment and Wage Statistics (OEWS) (}\url{https://www.bls.gov/emp/ind-occ-matrix/occupation.xlsx}).
Across all FOS, \hl{the decreasing values of KL divergence over time in Fig.{~\ref{fig:klDivergence_all_vs_csmath}}a indicate that the skills taught in course syllabi are becoming more similar to the skills required in the labor market. Specifically, comparisons between earlier syllabi (e.g., 00-03) and later labor force periods (e.g., 12-16) show a trend of decreasing divergence, reflecting an increasing alignment of skills.
\textit{This pattern indicates that educational entities are progressively aligning their course content more closely with the requirements of professional environments.} 
This alignment might result from a heightened recognition of occupational needs, enhancements in educational methodologies, or influences from regulatory agencies and corporate collaborations. Additionally, the reduction in KL divergence over successive periods underscores that recent syllabi not only integrate more pertinent skills but also likely eliminate outdated elements less relevant to contemporary professional demands. These gradual modifications suggest an encouraging evolution toward educational outputs that are directly advantageous for students transitioning into professional roles.}
These results suggest that taught skills are forward-looking.
However, going beyond existing research, our dataset enables us to make direct comparisons between specific FOS and labor market dynamics for individual occupations.
For example, motivated by earlier analysis of CS syllabi and CS-related job postings~\cite{borner2018skill}, we compare CS syllabi to Computer and Mathematical occupations (i.e., Standard Occupation Classification code: 15-0000. See Fig.~\ref{fig:klDivergence_all_vs_csmath}b). 
Despite the trend across all FOS, over time, the labor force skill distribution becomes increasingly dissimilar to older course syllabi which confirms the rapidly changing nature of such domains.
Comparing the KL Divergence scores of the syllabi among different periods (top left box of Fig.~\ref{fig:klDivergence_all_vs_csmath}b), we observe that syllabi are staying stagnant, and as a result, they are moving away from the frontier of knowledge required in the labor force.

\section*{Usage Notes}

Course-Skill Atlas offers a versatile tool for addressing a variety of research questions pertinent to education and workforce development across multiple domains. 
In the following, we \hl{briefly touch on potential research questions} utilizing this dataset, including exploring differences in skill sets across gender profiles in U.S. higher education, the trend of abilities in teaching activities, and  utilizing skill scores for salary estimation.
Lastly, we discuss our data's limitations.

\begin{itemize}

\item \hl{How does the specificity of skills taught in different college majors affect labor market outcomes, such as wages, career adaptability, and the likelihood of obtaining managerial positions?}

Research has consistently shown a strong relationship between college majors, skills, and wages~\cite{grogger1995changes,eide1994college,long2015completed}. For example, some majors may offer more diverse skill sets with more general skills profiles that lead to adaptable careers after a student graduates and enters the workforce~\cite{hemelt2023college}. The salary gap among majors is multifaceted, involving factors like labor market demands~\cite{altonji2016cashier} and major distribution's effect on gender wage disparities~\cite{eide1994college}.
\hl{There is a growing literature in labor and education economics on how general versus specific majors affect occupational choice and wages{~\cite{leighton2020labor, martin2022college}}. 
Majors with higher specificity, such as education and nursing, generally lead to higher earnings compared to more general majors like music and psychology, driven by higher hourly wages. 
However, graduates from specific majors are less likely to hold managerial positions, with those from majors of average specificity being most likely to become managers{~\cite{leighton2020labor}}.
Our dataset provides the opportunity to investigate such differences.}

\item \hl{How have teaching strategies and curriculum design evolved over time across different majors and universities?}

Our dataset enables the study of skill differences within and across majors and universities over time.
Taking active learning in social sciences as an example, a recent critique of active learning and the employability agenda in higher education within the social sciences~\cite{socialSciencesSkillsGap} identified an inadvertent neglect of key skills including reading, listening, and note-taking due to the lack of proper active learning strategies. 
\hl{The findings from such a direction of research could} pave the way for further investigation into how educational strategies can be developed to effectively balance traditional academic skills with the competencies essential for active learning environments.

    \item \hl{What role do educational institutions play in shaping the differences in skills between genders, particularly in relation to course syllabi?}

Existing research finds that males and females tend to possess different workplace skills on aggregate~\cite{christl2020gender,azmat2017gender}, which may correspond to gender stereotypes shaping careers~\cite{10.1093/ser/mwac062}.
But are these differences the result of education or labor market outcomes?
In general, these questions can only be studied through enrollment and graduation statistics from the US Department of Education without taking into account the granular differences in taught skills across different majors and institutions.
However, our dataset enables the study of this heterogeneity and, thus, creates an opportunity to explain career outcomes from the differences in skills taught during higher education---even differences within a given FOS based on varied enrollment across educational institutions.
\hl{Our dataset has the potential to explain the skill differences between majors and institutions based on course syllabi.}

\end{itemize}

\paragraph{Limitations}
This study produces a novel large-scale data set reflecting the skills taught to US college and university students across majors.
While useful for understanding one of the major pathways for workforce development in the US, there are some limitations to the current data set.
\hl{First, the syllabus dataset is, to our knowledge, the largest collection of university syllabi available, but as reported in{~\cite{biasi2022eduInov}} it is slightly skewed by school selectivity, overrepresenting Ivy-Plus, Elite, and selective public schools by 2.4 to 4.0 percentage points, while including less than 0.1 percent from non-selective institutions. 
Although the sample isn't biased by observable characteristics and consistently represents about $5\%$ of all courses taught in US institutions post 1998, the potential for unobservable selection remains a caveat in interpreting the results.}
Second, we propose a new approach for inferring taught skills (i.e., O\textsuperscript{*}NET DWAs and Tasks) from syllabus text, but it is difficult to confirm the effectiveness of our approach without wide-scale comprehensive exams to test the skills that students actually obtain during a course.
Such an effort would be extremely cumbersome because each student would ideally be assessed on over 2,000 DWAs; it's not clear how to empirically validate each of these DWAs and implementing such an examination across universities and majors throughout the US would be an immense undertaking.
Effectively, it is crucial to acknowledge that teaching does not necessarily equate to learning. 
\hl{Third, another limitation of this work relates to handling potential prerequisites. 
Prerequisites might appear in a syllabus in two forms.
If they appear as administrative details (e.g., the course code), since our data preparation pipeline removes such details, they will not affect the inferred skill vector.
On the other hand, when a syllabus includes the content of prerequisites --- similar to learning objectives, they are processed in the same manner as the skills for the course itself. 
However, due to the unstructured nature of each course description (i.e., presented in a single string format), we are unable to identify and exclude these prerequisites from the final skill scores. 
This limitation affects the accuracy of the skill assessment by conflating course skills with prerequisites.}
\hl{Fourth, due to the lack of enough metadata in the raw dataset, we are unable to distinguish between undergrad and graduate courses.}
Fifth, our approach relies on the O\textsuperscript{*}NET database which is designed to describe workers in the US workforce, and not explicitly designed to describe learning outcomes.
While O\textsuperscript{*}NET serves as a standardized taxonomy for communicating results to policymakers, its coverage across all occupations, and by extension, academic majors, is not uniformly comprehensive.
Sixth, existing research~\cite{collegeCoursework2019} show that the distribution of the course credits varies for college students even with the same field of study. 
Consequently, using field of study as a stand-in for an individual's complete set of skills is inadequate.
Due to the lack of data on enrollment per major and coursework taken by the students, we based our coverage analysis solely on the number of graduates per major.

\section*{Code availability}

The source code for \texttt{Syllabus2O\textsuperscript{*}NET} and \texttt{DWA2Ability} is available at \url{https://github.com/AlirezaJavadian/Syllabus-to-ONET}.

\section*{Acknowledgments}

This work received funding from Russell Sage Foundation (G-2109-33808) and is supported by the University of Pittsburgh Center for Research Computing.

\section*{Author contributions statement}

A.J.S. processed the data, performed all calculations, and produced all figures and statistics. 
All authors designed the research, analyzed the results, and reviewed the manuscript.

\section*{Competing interests} 
The authors declare no competing interests.

\section*{Acknowledgments}

This work received funding from Russell Sage Foundation and is supported by the University of Pittsburgh Center for Research Computing.

\section*{Author contributions statement}

A.J.S. processed the data, performed all calculations, and produced all figures and statistics. 
All authors designed the research, analyzed the results, and reviewed the manuscript.

\section*{Competing interests} 
The authors declare no competing interests.

\end{document}


\maketitle

\tableofcontents

\pagebreak

\section{Descriptive Statistics}

The Open Syllabus Project (OSP) Dataset\footnote{\url{https://opensyllabus.org/} (accessed
on 28\textsuperscript{th} November 2023)} is composed of nearly $8,000,000$ course syllabi worldwide among which $3,162,747$ syllabi across $62$ fields of study (FOS) belong to $2,761$ U.S. colleges and universities.
Tables~\ref{tab:fos_frequency} and~\ref{tab:year_frequency} list the frequency of syllabi per FOS and year respectively.
Table~\ref{tab:states_all_stats} details the geographical coverage of the OSP dataset.
Number of educational institutions within each state is obtained from the Carnegie Classification of Institutions of Higher Education (CCIHE).\footnote{\url{https://carnegieclassifications.acenet.edu/} (accessed
on 18\textsuperscript{th}Februaryy 2023)}
For example, Texas with $865,973$ syllabi has the largest number of syllabi ($27.85\%$) in the dataset.
According to CCIHE, there are $226$ universities and educational institutions located in Texas, among which $54.42\%$ have at least $8$ syllabi (25\textsuperscript{th} percentile) in the OSP dataset. 
Figure~\ref{fig:fos_per_year_frequency_heatmap} details the number of syllabi per FOS and year between 2000 and 2017 (see Figure~\ref{fig:two_digits_fos_yearly_cnt_2000_heatmap} for FOS based on 2-digit Classification of Instructional Programs (CIP)).
Figure~\ref{fig:Joint_Distribution_of_Syllabi_and_University_Counts} shows the joint distribution of the syllabus count and university count across all years and all FOS.
Table~\ref{tab:institution_frequency} lists the syllabus count per top 100 universities across all years and all FOS.

\input{figures/fos_frequency}
\input{figures/year_frequency}

\input{figures/states_all_stats}

\begin{figure}[H]
    \centering
    \includegraphics[trim=0 0 150 0,clip,width=1\textwidth]{figures/fos_yearly_cnt_2000_heatmap.pdf}
    \caption{Frequency of syllabi per field of study and year between 2000 and 2017.}
    \label{fig:fos_per_year_frequency_heatmap}
\end{figure}

\begin{figure}[H]
    \centering
    \includegraphics[trim=0 0 110 0,clip,width=1\textwidth]{figures/two_digits_fos_yearly_cnt_2000_heatmap.pdf}
    \caption{Frequency of syllabi per  field of study based on 2-digit CIP and year between 2000 and 2017.}
    \label{fig:two_digits_fos_yearly_cnt_2000_heatmap}
\end{figure}

\begin{figure}[H]
    \centering
    \includegraphics[trim=0 0 0 0,clip,width=0.40\textwidth]{figures/joint_dist_syll_univ_count.pdf}
    \caption{The syllabus count per university across all years and all FOS.}
    \label{fig:Joint_Distribution_of_Syllabi_and_University_Counts}
\end{figure}

\include{figures/institution_frequency}

\section{Skills Inference Framework (\texorpdfstring{\texttt{Syllabus2O\textsuperscript{*}NET})}{}}

Given the text from a course syllabus, we segment sentences using Stanza~\cite{qi2020stanza}, a tool designed to partition text into individual sentences. 
We extracted $322,473,524$ sentences from the course syllabi in our dataset. 
On average, each syllabus contains $101.96$ sentences (median $83$).
Each course syllabus starts as unstructured text lacking metadata to distinguish between course logistics (e.g., scheduling and grading rubrics) and learning outcomes (e.g., course content and learning objectives). 
Thus, we implement a human-in-the-loop approach to remove sentences pertaining to Course Logistics while keeping sentences about Learning Outcomes.
To do so, we compiled two distinct lists for labeling sentences. 
The list of Course Logistic related terms includes $356$ phrases that are common but unrelated to the course content (e.g., ``Plagiarism,'' ``Attendance,'' and ``Office hours''). 
The list of Learning Objective related terms includes $51$ phrases such as ``Analyze,'' ``Versus,'' and ``Outcome.'' 
The complete lists can be found on the code's GitHub page.\footnote{\url{https://github.com/AlirezaJavadian/Syllabus-to-ONET}}
We removed sentences from each syllabus that contained Course Logistic phrases or lacked Learning Objective phrases resulting in the removal of $85.82\%$ of the sentences.
Each syllabus contains nearly $17.61$ Learning Objective related sentences on average (median = $12$)
(see Table~\ref{tab:fos_sent_dist} for details on the statistics of Learning Objective sentence counts by FOS.).

\input{figures/fos_sent_dist}

\newpage

\section{Mapping \texorpdfstring{O\textsuperscript{*}NET DWAs to Abilities (\texttt{DWA2Ability})}:}

In the ``Gender, Education, and Skills'' and ``Temporal Trends in Teaching Activities'' sections, we use our data to explore results from the literature.
But existing studies are often forced to use less specific descriptions of skills, capabilities, or knowledge compared to the specificity of O\textsuperscript{*}NET DWAs and Tasks (see O\textsuperscript{*}NET's website for examples of DWAs\footnote{\url{https://www.onetcenter.org/dictionary/20.1/excel/dwa_reference.html} (accessed on 5\textsuperscript{th} February 2024)} and Tasks)\footnote{\url{https://www.onetcenter.org/dictionary/20.1/excel/task_statements.html} (accessed on 5\textsuperscript{th} February 2024)}.
We facilitate a fairer comparison between our data and existing research using the O\textsuperscript{*}NET Abilities taxonomy (see O\textsuperscript{*}NET's website \textit{Ability\footnote{\url{https://www.onetonline.org/find/descriptor/browse/1.A} (accessed on 5\textsuperscript{th} February 2024)}} for examples).
However, O\textsuperscript{*}NET does not provide a standardized cross-walk linking DWAs, tasks, and abilities. 
To bridge this gap, we introduce \texttt{DWA2Ability}.

We start with the O\textsuperscript{*}NET database\footnote{\url{https://www.onetonline.org/find/all} (accessed on 5\textsuperscript{th} February 2024)} profiles of DWAs for each occupation.
Next, we extract importance scores of each O\textsuperscript{*}NET ability
within each occupation.\footnote{\url{https://www.onetonline.org/find/descriptor/browse/1.A} (accessed on 5\textsuperscript{th} February 2024)} 
We formulate a map between the two sets of occupation profiles as a regression using DWAs as independent variables and ability scores as dependent variables. 
We train a Random Forest Regressor~\cite{scikit-learn} for each ability and fine-tune hyperparameters via Grid Search and 5-fold cross-validation. 
This approach yielded 52 models (i.e., one per O\textsuperscript{*}NET ability), each achieving mean squared error of at most 0.025 (see Table~\ref{tab:abilities_mse} for details on model performance). 
Using the trained models, we map syllabis' DWA scores to abilities.

\input{figures/abilities_mse}

\newpage

\section{Top 10 Detailed Work Activities (DWA) per Field of Study (FOS)}

Table~\ref{tab:top10DWAperFOS} lists the top 10 inferred workplace activities with the highest DWA scores per FOS. 
We mask the frequently occurring skills based on their prevalence. This was implemented by evaluating the top 100 DWAs for each FOS and masking those that were commonly observed. Note that masked DWAs remain accessible in the published dataset (see Section ``Skill Normalization'').

\input{figures/Tabletop10DWAperFOS}

\newpage

\section{Community Detection and Clustering}

We employ the Louvain community detection method~\cite{louvain} to cluster the most closely related academic majors. 
This process entailed averaging the scores for each DWA within identical majors, culminating in a unified vector representation for each field of study. 
This analysis discerned four distinct communities (see Figure~\ref{fig:DWA_fos_network}). 
The two most extensive communities generally correspond to Non-STEM (highlighted in green) and STEM (in blue) disciplines.
Notably, the smallest cluster, comprising Accounting, Economics, and Marketing, tends more towards Non-STEM fields, likely attributed to their social sciences orientation. 
The fourth community encapsulates fields such as Fitness, Education, and Nutrition.

Adopting an analogous approach with ``Tasks'' rather than DWAs, the results exhibit a significant variation (see Figure~\ref{fig:Task_fos_network}). 
The key divergence when employing ``Tasks'' is the formation of three communities instead of four. 
The group previously including Accounting, Economics, and Marketing amalgamated with the broader STEM category, indicating a shift in community dynamics based on the classification criterion used.

\begin{figure}[H]
    \centering
    \includegraphics[trim=220 320 190 320,clip,width=1\textwidth]{figures/DWA_fos_network.pdf}
    \caption{\textbf{The similarity of FOS according to their DWA profiles.} The graph represents the results of the Louvain community detection method~\cite{louvain} to identify the similar FOS. (\textit{Note:} Node size represents the Degree Centrality).}
    \label{fig:DWA_fos_network}
\end{figure}

\begin{figure}[H]
    \centering
    \includegraphics[trim=220 320 170 310,clip,width=1\textwidth]{figures/Task_fos_network.pdf}
    \caption{\textbf{The similarity of FOS according to their Task profiles.} The graph represents the results of the Louvain community detection method~\cite{louvain} to identify the similar FOS. (\textit{Note:} Node size represents the Degree Centrality)}
    \label{fig:Task_fos_network}
\end{figure}

Moreover, we employ agglomerative hierarchical clustering technique~\cite{johnson1967hierarchical} on the DWA vector representations of academic majors, aiming to elucidate their hierarchical relationships. 
Hierarchical clustering generates a nested sequence of clusters, allowing for an in-depth exploration of clusters at varying levels of granularity without predefining a specific number of categories (in this context, groups of majors). 
In this framework, FOS are deemed similar if they share akin work activities (see Figure~\ref{fig:DWA_field_name_dendogram} for DWAs and Figure~\ref{fig:Task_field_name_dendogram} for Tasks). 

The resulting dendrogram illustrates that majors cluster in a manner consistent with expectations. 
For instance, in case of DWAs (see Figure~\ref{fig:DWA_field_name_dendogram}), Marketing and Accounting emerge as closely related, akin to the pairing of Linguistics and History. Progressing through the hierarchy, these clusters show increasing similarity to other fields, such as Anthropology. 
Notably, just prior to the final clustering step, which amalgamates all majors (indicated in blue), two predominant clusters are discernible: one representing STEM majors (in green) and the other Non-STEM majors (in orange).

\begin{figure}[H]
    \centering
    \includegraphics[trim=10 100 150 100,clip,width=0.82\textwidth]{figures/DWA_field_name_dendogram.pdf}
    \caption{\textbf{The similarity of FOS according to their DWA profiles.} The heatmap shows the Spearman’s rank correlation between the fields of study and the dendrogram represents the results of hierarchically clustering similar FOS.}
    \label{fig:DWA_field_name_dendogram}
\end{figure}

\begin{figure}[H]
    \centering
    \includegraphics[trim=10 100 150 100,clip,width=0.82\textwidth]{figures/Task_field_name_dendogram.pdf}
    \caption{\textbf{The similarity of FOS according to their Task profiles.} The heatmap shows the Spearman’s rank correlation between the fields of study and the dendrogram represents the results of hierarchically clustering similar FOS.}
    \label{fig:Task_field_name_dendogram}
\end{figure}

\section{Gender, Education, and Skills}

To make a more direct comparison with the OECD report, we quantify these differences at the O\textsuperscript{*}NET Ability level using our \texttt{DWA2Ability} trained models (see Section ``Mapping O\textsuperscript{*}NET DWAs to Abilities (\texttt{DWA2Ability})''). 
In Table~\ref{tab:abilities_gender_proportion}, we compare the percentage of the difference between female students' abilities and male students' abilities

\input{figures/abilities_gender_proportion}

\newpage

\section{Taught Skill Distinctiveness and Labor Market Outcomes}

We compare the distinctiveness of FOS's taught skills with  earnings of 25- to 29-year-old bachelor’s degree holders reported by the National Center for Education Statistics (NCES)\footnote{\url{https://nces.ed.gov/programs/digest/d19/tables/dt19_505.10.asp} (accessed on 12\textsuperscript{th} February 2024)} (see Table~\ref{tab:salary}).
We measure skill distinctiveness by noting various percentile skill RCA score within each FOS (see figures~\ref{fig:rca_vs_salary_60}, \ref{fig:rca_vs_salary_70}, \ref{fig:rca_vs_salary_75}, \ref{fig:rca_vs_salary_80}, and \ref{fig:rca_vs_salary_90}); under this measure, FOS with higher diversity scores teach more distinctive skills compared to other FOS.

\input{figures/salary}

\begin{figure}[H]
    \centering
    \includegraphics[trim=30 0 30 50,clip,width=1\textwidth]{figures/rca_vs_salary_60.pdf}
    \caption{\textbf{Skill distinctiveness versus salary.} The x-axis shows the skill distinctiveness as RCA value at the 60\textsuperscript{th} percentile within each field of study and the y-axis shows the corresponding median annual earnings (salary) obtained from NCES. The line represents a regression line with $R^2$ = 0.006 (\textit{p-value} = 0.667).}
    \label{fig:rca_vs_salary_60}
\end{figure}

\begin{figure}[H]
    \centering
    \includegraphics[trim=30 0 30 50,clip,width=1\textwidth]{figures/rca_vs_salary_70.pdf}
    \caption{\textbf{Skill distinctiveness versus salary.} The x-axis shows the skill distinctiveness as RCA value at the 70\textsuperscript{th} percentile within each field of study and the y-axis shows the corresponding median annual earnings (salary) obtained from NCES. The line represents a regression line with $R^2$ = 0.154 (\textit{p-value} = 0.026).}
    \label{fig:rca_vs_salary_70}
\end{figure}

\begin{figure}[H]
    \centering
    \includegraphics[trim=30 0 30 50,clip,width=1\textwidth]{figures/rca_vs_salary_75.pdf}
    \caption{\textbf{Skill distinctiveness versus salary.} The x-axis shows the skill distinctiveness as RCA value at the 75\textsuperscript{th} percentile within each field of study and the y-axis shows the corresponding median annual earnings (salary) obtained from NCES. The line represents a regression line with $R^2$ = 0.250 (\textit{p-value} = 0.004).}
    \label{fig:rca_vs_salary_75}
\end{figure}

\begin{figure}[H]
    \centering
    \includegraphics[trim=30 0 30 50,clip,width=1\textwidth]{figures/rca_vs_salary_80.pdf}
    \caption{\textbf{Skill distinctiveness versus salary.} The x-axis shows the skill distinctiveness as RCA value at the 80\textsuperscript{th} percentile within each field of study and the y-axis shows the corresponding median annual earnings (salary) obtained from NCES. The line represents a regression line with $R^2$ = 0.290 (\textit{p-value} = 0.001).}
    \label{fig:rca_vs_salary_80}
\end{figure}

\begin{figure}[H]
    \centering
    \includegraphics[trim=30 0 30 50,clip,width=1\textwidth]{figures/rca_vs_salary_90.pdf}
    \caption{\textbf{Skill distinctiveness versus salary.} The x-axis shows the skill distinctiveness as RCA value at the 90\textsuperscript{th} percentile within each field of study and the y-axis shows the corresponding median annual earnings (salary) obtained from NCES. The line represents a regression line with $R^2$ = 0.294 (\textit{p-value} = 0.001).}
    \label{fig:rca_vs_salary_90}
\end{figure}